\documentclass[11pt,a4paper]{article}
\usepackage{jheppub}
\usepackage{graphicx}
\usepackage{amssymb}
\usepackage{amsmath}
\usepackage{latexsym}

\newcommand{\de}{\partial} 
\newcommand{\Braket}[3]{\langle #1 \lvert #3 \rvert #2 \rangle }  
\newcommand{\Dslash}{D \! \! \!  \!  /} 
\newcommand{\slashed}[1]{#1 \!\!\!\! / \,\,}
\newcommand{\media}[1]{\langle #1 \rangle} 
\newcommand{\D}{\mathcal{D}}
\newcommand{\be}{\begin{equation}}
\newcommand{\ee}{\end{equation}}
\newcommand{\bea}{\begin{eqnarray}}
\newcommand{\eea}{\end{eqnarray}}

\title{Quantum theory of massless $\bf (p,0)$-forms}
\author{Fiorenzo Bastianelli}
\author{and Roberto Bonezzi}
\affiliation{Dipartimento  di Fisica, Universit{\`a} di Bologna and\\
INFN, Sezione di Bologna, via Irnerio 46, I-40126 Bologna, Italy}

\emailAdd{bastianelli@bo.infn.it}\emailAdd{bonezzi@bo.infn.it}

\abstract{We describe the quantum theory of massless $(p,0)$-forms that satisfy  a suitable holomorphic generalization
of the free Maxwell equations on K\"ahler spaces. These equations arise by first-quantizing a  spinning
particle with a U(1)-extended local supersymmetry on the worldline.
Dirac quantization  of the spinning particle produces a physical Hilbert space made up of  $(p,0)$-forms that
satisfy holomorphic Maxwell equations coupled to the background K\"ahler geometry, containing in particular
a charge that measures the amount of coupling to the U(1) part of the U($d$) holonomy group of the
$d$-dimensional K\"ahler space. The relevant differential operators appearing in these equations are a
twisted exterior holomorphic derivative $\partial_q$ and its hermitian conjugate $\partial_q^\dagger$
(twisted Dolbeault operators with charge $q$). The particle model is used to obtain a worldline representation of the
one-loop effective action of the $(p,0)$-forms. This representation allows to compute the first
few heat kernel coefficients contained in the local expansion of the effective action and to
derive duality relations between $(p,0)$ and $(d-p-2,0)$-forms
that include a topological mismatch appearing at one-loop.}


\begin{document}
\maketitle
\flushbottom

\section{Introduction}
In this paper we wish to describe the quantization of $(p,0)$-form gauge fields $A$,
defined on K\"ahler spaces, which satisfy a holomorphic generalizations of the free Maxwell
equations
\be  \label{maxeq}
\partial^\dagger_q F =0
\;, \qquad
F=  \partial_q A
\ee
where the twisted exterior holomorphic derivative $\partial_q = \partial + q\Gamma $
contains a coupling to the U(1) part of the U($d$) holonomy group of the $d$-dimensional K\"ahler space
($\Gamma \equiv \Gamma_\mu dx^\mu = \Gamma_{\mu\nu}^\nu dx^\mu$
with $x^\mu$ complex coordinates) and is  a nilpotent operator ($\partial_q^2=0$).
It is the natural generalization on K\"ahler manifolds of the standard quantum theory of
differential $p$-forms  $A$ that satisfy the Maxwell equation $d^\dagger d A =0$
and enjoy a gauge invariance of the form $\delta A = d \lambda$ where $\lambda$ is a  $(p-1)$-form.

We are going to use a worldline approach in which the physical degrees of freedom
of the $(p,0)$ gauge field are carried by a spinning particle with a U(1)-extended local supersymmetry
on the worldline. This approach parallels the one used in \cite{Bastianelli:2005vk, Bastianelli:2005uy}
for standard differential $p$-forms, which allowed  to derive quite elegantly  exact duality relations,
compute heat kernel coefficients, and calculate the one-loop contribution
to the graviton self-energy (the  two-point function of the $p$-form stress tensors).
In that case, some of those results had already been obtained previously using standard QFT tools,
which include the correct way of covariantly gauge fixing the $p$-form gauge symmetries
\cite{Siegel:1980jj,Thierry-Mieg:1980it}, and the derivation of topological mismatches between the
unregulated effective actions of dual forms \cite{Duff:1980qv,Schwarz:1984wk}.
In the present case we proceed directly by employing a worldline representation, and use it
to study the one-loop effective action as function of the background metric,
compute the first few heat kernel coefficients that characterize it, and derive exact duality relations
between the effective actions of  $(p,0)$ and $(d-p-2,0)$-forms.

The spinning particle that we use to treat the $(p,0)$-form gauge fields is a
U(1) spinning particle, by which we mean a particle model that contains a U(1)-extended local supersymmetry
on the worldline. The corresponding supersymmetry charges $Q$ and $\bar Q$  are realized quantum
mechanically by twisted Dolbeault operators $\partial_q$ and $\partial_q^\dagger$ acting on the Hilbert space
of  the $(p,0)$-forms with any allowed $p$. This model was derived sometimes ago in  \cite{Marcus:1994em} to describe
the so-called topological B model in a  simple setting, and then generalized in  \cite{Marcus:1994mm} to a class of
U($N$) spinning particles  that have been used in \cite{Bastianelli:2009vj} to derive higher spin equations
on complex manifolds. For $N=1$ those equations reduce precisely to the ones that are analyzed in the present paper.
At the ungauged level, i.e. when supersymmetry is kept only as a rigid symmetry,  one obtains
a related sigma model that has been used recently in \cite{Ivanov:2010ki,Smilga:2011wi} to study the twisted Dolbeault
complex and related index theorems.

We present our material in the following way. We start with section 2 describing the canonical quantization
of the U(1) spinning particle in flat complex space. This allows to introduce in a simple context the holomorphic
equations briefly presented above.
In section 3 we consider a generic K\"ahler space, and discuss canonical
quantization of the spinning particle, paying attention to the ordering ambiguities that allow the introduction
of a free coupling to the U(1) part of the holonomy group of the background K\"ahler space.
We describe how the supercharges of the model realize the twisted Dolbeault operators $\partial_q$
and $\partial_q^\dagger$, with $q$ indicating the free coupling constant just mentioned.
Then we consider the ungauged model (rigid susy), and use operatorial methods to compute perturbatively
the transition amplitude as well as  path integral methods to obtain the Dirac index $(q=\frac14$)
and its twisted versions $(q\neq\frac14$).
In section 4 we consider the gauged model, i.e. the complete U(1) spinning particle, to give a worldline representation
of the one-loop effective action of the $(p,0)$-form gauge fields, and use it to compute the first
few heat kernel coefficients characterizing the effective action.
This provides the quantization with worldline methods of the gauge invariant field equations
$\partial^\dagger_q \partial_q A =0$.
As a side result, we present the heat kernel coefficients for the ungauged model as well, that corresponds to the
worldline quantization of the field equations $(\partial^\dagger_q \partial_q +\partial_q \partial^\dagger_q ) B =0$,
which do not carry any gauge invariance.
In section 5 we discuss various dualities and derive topological mismatches appearing at one-loop,
checking them versus the explicit results found in the preceding section.
Presenting the (unregulated) effective action of a $(p,0)$-form gauge field with a U(1) charge $q$ in the form
of an integral over proper time of a corresponding density,
$Z_p(q)=\int \frac{d\beta}{\beta} \mathcal{Z}_p(q, \beta)$, we find
a duality between a $(p,0)$-form and a $(d-p-2,0)$-form described  by the following relation
 \begin{equation}
\mathcal{Z}_{p}(q,\beta) =\mathcal{Z}_{d-p-2}(\tfrac12-q,\beta)
+ (-1)^{p} \mathcal{Z}_{d-1}(\tfrac12-q,\beta)
+ (-1)^{p} (p+1) \; {\rm ind} (\Dslash_{q-1/4})
\end{equation}
 where
$\mathcal{Z}_{d-1}(\tfrac12-q,\beta)$ is a purely topological contribution (no propagating degrees of freedom are associated to
a $(d-1,0)$-form for $d>1$) that can be related to the analytic torsion,
and ${\rm ind} (\Dslash_{q-1/4})$ is the index of the (twisted for $q\neq \tfrac14$) Dirac operator.
Finally, we present our conclusions and perspectives  in section 6.

\section{Free particle and canonical quantization}
In this section we review the free U(1) spinning particle and its Dirac quantization to
describe in the simple context of $\mathbb{C}^d$, the  flat complex space,
how the Maxwell equations for a $(p,0)$-form emerge naturally from first-quantization.
The particle system of interest is constructed by first considering
a supersymmetric particle that  produces a Hilbert space ${\cal H}$
formed by the sum of all $(p,0)$-forms with any allowed $p$,
\begin{equation}
{\cal H}=\bigoplus_{p=0}^d \Lambda^{p,0}(\mathbb{C}^d)
\end{equation} 
where $\Lambda^{p,q}$ indicates the space of $(p,q)$-forms.
This mechanical model contains conserved supercharges $Q$ and $\bar Q$
that are realized on the Hilbert space by the Dolbeault operator
$\partial$ and its hermitian conjugate $\partial^\dagger$. It is seen that
the supercharges belong to a multiplet of conserved
charges containing the hamiltonian $H$ and a U(1) charge $J$ as well. Altogether
these charges satisfy a U(1)-extended supersymmetry algebra.
Gauging all of them produces the action of the U(1) spinning particle
that leads to the  quantum theory of a $(p,0)$-form
obeying the Maxwell equations in (\ref{maxeq}).
The details are as follows.

We consider a particle moving in flat complex space $\mathbb{C}^d$, described by the  complex coordinates
 ($x^\mu, \bar{x}^{\bar{\mu}}$) with $\mu=1,..,d$. The particle carries additional
 degrees of freedom associated to the Grassmann variable $\psi^\mu$
 and its complex conjugate $\bar{\psi}^{\bar{\mu}}$. Indices are lowered and raised with the flat metric
$\delta_{\mu\bar{\nu}}$ and its inverse.\footnote{
We often use a redundant notation by indicating complex conjugate variables by using a bar on both
the variable itself and its indices, such as $\bar{x}^{\bar{\mu}}$, $\bar{p}_{\bar{\mu}}$ or
 $\bar \partial_{\bar{\mu}}$. This allows for a quick interpretation of various formulas, containing  for example
  $\bar p^\mu= g^{\mu\bar\nu}\bar p_{\bar \nu}$ or similar tensors with upper indices.
 No confusion should arise whenever we use such a redundant notation.}
With these ingredients,
the ungauged model is identified by the phase space action
\begin{equation}
S=\int dt \big[p_\mu\dot{x}^\mu+\bar{p}_{\bar{\mu}}\dot{\bar{x}}^{\bar{\mu}}
+i\bar{\psi}_\mu\dot{\psi}^\mu
-p_\mu\bar{p}^\mu
\big]
\end{equation}
that indeed describes a free motion on  $\mathbb{C}^d$ .
The conserved charges
\begin{equation} \label{charges}
H=p_\mu \bar{p}^\mu \ , \quad
Q=\psi^\mu p_\mu  \ , \quad
\bar{Q}=\bar{\psi}^{\bar{\mu}}\bar{p}_{\bar{\mu}} \ , \quad
J=\psi^\mu\bar{\psi}_\mu
\end{equation}
guarantee the existence of a U(1)-extended supersymmetry algebra on the worldline.
Canonical quantization shows immediately that the corresponding Hilbert space
can be realized by the set of all $(p,0)$-forms with $p=0,1,...,d$. In fact, the
elementary commutation relations obtained from the classical Poisson brackets read
\begin{equation} \label{elementary comm relations}
[x^\mu, p_\nu] = i\delta^\mu_\nu \;, \qquad
[\bar x^{\bar \mu}, \bar p_{\bar \nu} ]=i\delta^{\bar\mu}_{\bar \nu}\ , \qquad
\{\psi ^\mu,\bar\psi_\nu \} =  \delta^\mu_\nu \;.
\end{equation}
By considering $(x^\mu, \bar x^{\bar \mu},\psi^\mu)$ as coordinates and
$(p_\mu, \bar p_{\bar \mu}, \bar\psi_\mu)$ as momenta, one may realize the latter as
differential operators with respect to the former,
\begin{equation} 
p_\mu =-i\partial_\mu \;,\qquad
\bar p_{\bar \mu}=-i\bar \partial_{\bar \mu} \;, \qquad
\bar\psi_\mu = \frac{\partial}{\partial \psi^\mu}
\end{equation}
(we use left derivative for Grassmann variables), so that a generic wave function
$\phi(x,\bar x, \psi)$
has a finite expansion with respect to the Grassmann variables $\psi^\mu$,
and contains all possible differential $(p,0)$-forms up to $p=d$
\begin{equation} \label{wave function}
\phi(x,\bar x, \psi)= F(x,\bar x) + F_\mu(x,\bar x)\psi^\mu
+\frac12 F_{\mu\nu}(x,\bar x)\psi^\mu\psi^\nu
+ ...+\frac{1}{d!}F_{\mu_1..\mu_d}(x,\bar x)\psi^{\mu_1}..\psi^{\mu_d}\;.
\end{equation}
There are a total of $2^d$ independent components, which equals the number
of the independent components of a Dirac fermion. This is not a coincidence, as it is known that
on K\"ahler manifolds the space of all $(p,0)$-forms is equivalent to the Hilbert space of a Dirac fermion,
see appendix \ref{app:Dirac}.
The Hilbert space metric is the one that emerges naturally by considering
coherent states for worldline fermions, and takes the following schematic form
\begin{equation} \label{Hmetric}
\langle \chi|\phi \rangle = \int dx d\bar x d \psi d\bar \psi\;  {\rm e}^{\bar \psi \psi}\;
\overline{\chi(x,\bar x, \psi)}\; \phi(x,\bar x, \psi)
\end{equation}
so that $\bar{x}^{\bar{\mu}}$ is the hermitian conjugate of $x^\mu$,
$\bar{p}_{\bar{\mu}}$ is the hermitian conjugate of $p_\mu$,
and $\bar{\psi}^{\bar{\mu}}$ is the hermitian conjugate of $\psi^\mu$
(note that in flat space $\bar{\psi}_{\mu}=\bar{\psi}^{\bar{\mu}}$).

On the Hilbert space thus constructed the quantized conserved charges are represented
by differential operators. In particular,
the operator $iQ=\psi^\mu \partial_\mu$ naturally
acts as the Dolbeault operator $\partial=dx^\mu \wedge \partial_\mu $
on $(p,0)$-forms. Similarly
$i\bar Q = \bar\partial^\mu \frac{\partial}{\partial \psi^\mu}$
corresponds, up to a sign,  to its adjoint $\partial^\dagger$
acting on $(p,0)$-forms.
The Hamiltonian is given by the laplacian $ H=-\bar \partial^\mu \partial_\mu$.
Finally, the U(1) charge operator $J=\psi^\mu \frac{\partial}{\partial \psi^\mu}$
counts the rank $p$ of a $(p,0)$-form, up to a normal ordering ambiguity
that we shall discuss in a moment.
The $U(1)$-extended supersymmetry algebra satisfied by these operators is easily
computed and reads
\begin{equation}\label{algebra}
\{Q,\bar Q\} =  H \;,\qquad
[J,Q] = Q \;,\qquad
[J,\bar Q] = - \bar Q
\end{equation}
while other (anti)-commutators vanish.

The U(1) spinning particle we shall consider is obtained by gauging all of the symmetries generated
by the charges in (\ref{charges}). The emerging model has a U(1)-extended local
supersymmetry on the worldline, and it is characterized by the phase space action
\begin{equation} \label{spinning particle}
S=\int dt \Big[p_\mu\dot x^\mu+\bar p_{\bar\mu}\dot{\bar x}^{\bar\mu}+i\bar\psi_\mu\dot\psi^\mu-eH-i\chi\bar Q-i\bar\chi Q+a(J-s) \Big]
\end{equation}
where $G\equiv (e,\chi,\bar\chi,a)$ are the worldline gauge fields that make local
the symmetries generated by the constraints $T\equiv (H, Q, \bar Q, J-s)$.
The coupling $s$ in (\ref{spinning particle}) is a Chern-Simons coupling
(note that its redefinition can take into account different ordering prescriptions
that may be chosen when constructing the operator $J$ in canonical quantization).
It is crucial for obtaining quantum mechanically a non-empty model, and for this purpose
it must be quantized to integer values.
In a Dirac quantization scheme, one can gauge-fix the worldline gauge fields to predetermined values,
and require the constraints to annihilate physical states: $T|\phi_{phys}\rangle=0$.
The constraint $J-s=0$ selects $(s,0)$-forms
\begin{equation}
 \phi_{phys}(x,\bar x, \psi) =
\frac{1}{s!}F_{\mu_1..\mu_s}(x,\bar x)\psi^{\mu_1}..\psi^{\mu_s}
\end{equation}
so that the model may be  non-empty if the coupling $s$ in an integer with values $0\leq s\leq d$.
For convenience we set $s \equiv p+1$, so that the $J$ constraint selects the $(p+1,0)$-form $F_{(p+1,0)}$
containing $p+1$ holomorphic lower indices.
Then the $Q$ constraints $Q |\phi_{phys}\rangle=0$ is equivalent to
\be
\partial F_{(p+1,0)} =0
\ee
which can be solved by $F_{(p+1,0)} = \partial A_{(p,0)}$ up to a gauge transformation
$\delta A_{(p,0)} = \partial \lambda_{(p-1,0)}$.
Finally, the $\bar Q$ constraint gives the remaining Maxwell equation
\be
\partial^\dagger F_{(p+1,0)} =0
\ee
that reads as $\partial^\dagger \partial   A_{(p,0)}=0$ in terms of the gauge potential.

In components, the equations of motion of the field strength take the form
\begin{equation}
\partial_{[\mu} F_{\mu_1..\mu_{p+1}]} =0 \;, \qquad \bar \partial^{\mu_1}F_{\mu_1..\mu_{p+1}} =0 
\end{equation}
and are expressed in terms of the gauge potential by
\begin{equation}\label{A equation flat}
\begin{split}
& F_{\mu_1..\mu_{p+1}} =\partial_{\mu_1} A_{\mu_2..\mu_{p+1}} \pm {\rm cyclic\ permutations} \\[2mm]
&\bar \partial^\mu \partial_\mu A_{\mu_1..\mu_p} +(-1)^p p\:\bar \partial^\mu
\partial_{[\mu_1} A_{\mu_2..\mu_p]\mu}=0
\end{split}
\end{equation}
with square brackets indicating weighted antisymmetrization. These equations are
invariant under the gauge transformations $\delta A_{(p,0)} = \partial \lambda_{(p-1,0)}$, i.e.
\begin{equation}
\delta A_{\mu_1..\mu_p} =\partial_{\mu_1} \lambda_{\mu_2..\mu_{p}} \pm {\rm cyclic\ permutations} \;.
\end{equation}
In particular, for $p=1$ one obtains the simple holomorphic Maxwell equations
\begin{equation}
\bar \partial^\mu F_{\mu\nu} =0 \;, \qquad F_{\mu\nu} =\partial_\mu A_\nu - \partial_\nu A_\mu
\end{equation}
with gauge symmetry $\delta A_\mu =\partial_\mu\lambda$.

Of course, different models can be obtained by gauging different subgroups of the
U(1) extended supermultiplet of charges. In particular, if one decides to gauge only the hamiltonian $H$
and the real linear combination of the supercharges $Q+\bar Q$,
one obtains a first quantized description of a massless Dirac field. In fact,
on K\"ahler manifolds the Hilbert space of a fermion
corresponds to the collection of all $(p,0)$-forms, and the
Dirac operator corresponds to  the real supercharge $Q+\bar Q\sim \partial+\partial^\dagger$
(although on curved K\"ahler manifolds this happens only  when the Dolbeault operator acquires a specific coupling
to the U(1) part of the holonomy group, as discussed in appendix \ref{app:Dirac}).
Thus, a massless Dirac field in first quantization is obtained by quantizing the worldline
action
\begin{equation} \label{spinning particle:Dirac}
S=\int dt \Big[p_\mu\dot x^\mu+\bar p_{\bar\mu}\dot{\bar x}^{\bar\mu}+i\bar\psi_\mu\dot\psi^\mu-eH
-i\chi (Q + \bar Q) \Big]
\end{equation}
where $\chi$ is a real worldline gravitino.

\section{Coupling to gravity, transition amplitude, and the Dirac index}
\label{sec-3}

We are now going to consider the coupling to an arbitrary background K\"ahler metric.
It is useful to start with the ungauged version of the particle, which
provides us with a nonlinear sigma model that contains already all operators
of interest. As a preparation for subsequent applications, we
evaluate its quantum mechanical transition amplitude and compute the Dirac index
by considering its partition function with periodic boundary conditions.
The notations employed are listed in appendix \ref{app:notations}.

A simple way to introduce couplings to
the background K\"ahler metric, while maintaining the U(1)-extended supersymmetry, is to consider
the covariantization of the symmetry charges $J, Q, \bar Q$, and then
imposing the susy algebra to obtain the correct hamiltonian $H$.
We consider the Grassmann variables  $\psi^\mu$ and $\bar\psi_\mu$ as tensors transforming
 under holomorphic change of coordinates according to the position of their indices.
Then the classical charge
$J_{cl}=\psi^\mu\bar\psi_\mu $ is already covariant (a scalar). As for the susy charges, it is convenient to substitute
the momenta $(p_\mu, \bar{p}_{\bar\mu})$ there contained  by ``covariant" momenta
$(\pi_\mu, \bar{\pi}_{\bar\mu})$ defined by
\begin{equation}
\pi_\mu=p_\mu+i\Gamma^\lambda_{\mu\nu}\psi^\nu \bar\psi_\lambda\;,
\qquad
\bar\pi_{\bar\mu}=\bar p_{\bar\mu}
\end{equation}
that indeed are characterized by a Poisson bracket  proportional to the curvature tensor
\begin{equation}
\{\pi_\mu, \bar\pi_{\bar\nu}\}_{_{PB}}= i R_{ \mu{\bar \nu}}{}^\lambda{}_\sigma  \psi^\sigma \bar\psi_\lambda\;.
\end{equation}
Thus one obtains
\begin{equation} \label{3.1}
\begin{split}
& Q_{cl}=\psi^\mu \pi_\mu
= \psi^\mu\big(p_\mu+i\Gamma^\lambda_{\mu\nu}\psi^\nu\bar\psi_\lambda\big)
=\psi^\mu p_\mu \\[2mm]
& \bar Q_{cl}=\bar\psi_\mu g^{\mu\bar\nu}\bar \pi_{\bar\nu}
= \bar\psi_\mu g^{\mu\bar\nu}\bar p_{\bar\nu} \;.
\end{split}
\end{equation}
Thanks to the anticommuting character of the Grassmann variables,
the term with the Christoffel connection vanishes in $Q_{cl}$, and the curved K\"ahler
metric appears only in $\bar Q_{cl}$.
Now one can compute their Poisson bracket, and check that the
U(1)-extended supersymmetry algebra is realized with the classical hamiltonian
\begin{equation} \label{3.2}
 H_{cl} = g^{\mu\bar\nu}\bar p_{\bar\nu}\big(p_\mu+i\Gamma^\lambda_{\mu\sigma}\psi^\sigma\bar\psi_\lambda\big)
\;.
\end{equation}
With this $H_{cl}$ the phase space action for the searched for covariant model reads
\begin{equation} \label{3.3}
S_{ph} = \int dt\,\Big[p_\mu\dot x^\mu+\bar p_{\bar\mu}\dot{\bar
x}^{\bar\mu}+i\bar\psi_\mu\dot\psi^\mu-H_{cl}\Big] \;.
\end{equation}
Eliminating the momenta $(p,\bar p)$ one obtains the corresponding
nonlinear sigma model in configuration space
\begin{equation} \label{ungauged}
S_{con} = \int dt\,\Big[g_{\mu\bar\nu}\dot x^\mu \dot{\bar x}^{\bar\nu}
+i\bar\psi_\mu D_t\psi^\mu\Big]
\end{equation}
where the covariant time derivative is given by $D_t\psi^\mu=\dot\psi^\mu+\dot x^\nu\Gamma^\mu_{\nu\lambda}\psi^\lambda$.
This action is real up to boundary terms.
Of course, one could have proceeded differently, covariantizing the configuration space action first and casting it
in hamiltonian form afterwards.

Now, we may study canonical quantization.
As outlined in the flat space case, canonical quantization produces an Hilbert space
formed by the space of all $(p,0)$-forms living on the K\"ahler manifold $M$, that is
${\cal H}=\bigoplus_{p=0}^d \Lambda^{p,0}(M)$. One may again expect the susy charges $Q$ and $\bar Q$
to be represented by the Dolbeault operators $\partial$ and $\partial^\dagger$, and the real charge $Q+\bar Q$
by the Dirac operator ${\slashed D}=\gamma^\mu D_\mu + \gamma^{\bar \mu} D_{\bar\mu}$.
This is correct on manifolds of SU($d$) holonomy, where the Dirac operator indeed satisfies
${\slashed D}=\gamma^\mu D_\mu + \gamma^{\bar \mu} D_{\bar\mu}\sim \partial+\partial^\dagger$.
However, on generic K\"ahler manifolds of U($d$)  holonomy, one finds a nontrivial coupling
to the U(1) part of the U($d$)=U(1)$\times$ SU($d$) connection. This is required by the couplings of the
Dirac operator, see appendix \ref{app:Dirac}.
Therefore, let us analyze in more details
the operatorial realization of the susy charges in terms of differential  operators to appreciate
how the ordering ambiguities leave enough room for the emergence of an additional free coupling
to the U(1) part of the connection. This coupling is fixed if one wants to reproduce the Dirac operator, 
otherwise it can be considered arbitrary if one wishes to consider more general (covariant) models.

The commutation relations  between the basic dynamical variables are as in
 (\ref{elementary comm relations}), however the construction of composite operators may suffer from ordering ambiguites.
 The latter can  be resolved  partially by  (\emph{i})
 requiring covariance under holomorphic change of coordinates and  (\emph{ii}) imposing the
  correct hermiticity properties that arise from the analogous properties under complex conjugation
  of the classical model.
    As we shall see this leaves the possibility of having a free U(1) charge
     in the quantum model.  Generically on K\"ahler manifolds  there is no need to introduce flat indices,
 and we will proceed that way as much as we can.
 The U(1) R-charge $J$ is quadratic, and suffers only of a quite mild ordering ambiguity upon quantization.
Having in mind path integral calculations, where ordering ambiguities take the form of
different regularizations of the path integral, we choose an ordering that is naturally related to the way we regulate
and compute the path integral. This corresponds to the antisymmetrization of the quadratic fermionic term
\begin{equation}
J_{cl} =\psi^\mu\bar\psi_\mu \quad \to\quad J=
 \frac12 (\psi^\mu\bar \psi_\mu -\bar \psi_\mu\psi^\mu)= \psi^\mu\bar \psi_\mu -\frac{d}{2}\;.
  \end{equation}
 As already mentioned, different orderings can be taken into account by a redefinition of the
Chern-Simons coupling of the U(1) spinning particle. In particular, choosing
the value $s \equiv p+1-\frac{d}{2}$ in the covariant version of (\ref{spinning particle})
(so that $J-s=\psi^\mu\bar\psi_\mu -(p+1)$ as an operator)
allows to project onto the sector of the Hilbert space
containing $(p+1,0)$-forms only.
The covariance of this operator is manifest.

A bit more subtle is the construction of the covariant supercharges. It is useful to start again from covariant
momenta, as past experience with the standard spinning particle on riemannian manifolds indicates.
In this case (as opposite to the riemannian case) covariance
is not enough to fix all ordering ambiguities, and one finds a nontrivial coupling to the U(1) part of the holonomy
\begin{equation}
\begin{split}
&\pi_\mu=p_\mu+i\Gamma^\lambda_{\mu\nu}\psi^\nu\bar\psi_\lambda
\qquad  \to \qquad
\pi_\mu=p_\mu+i\Gamma^\lambda_{\mu\nu} \psi^\nu\bar\psi_\lambda -i q \Gamma_\mu \\[1mm]
&\bar\pi_{\bar\mu}=\bar p_{\bar\mu} \qquad  \qquad  \qquad  \quad  \
\to \qquad \bar\pi_{\bar\mu}=\bar p_{\bar\mu} +i q \Bar \Gamma_{\bar \mu}
\end{split}
\end{equation}
where on the left hand side we have listed the classical expressions, and on the right hand side
the quantum expressions.
A different ordering of the term with the fermionic operators can be compensated by a redefinition
of the charge $q$. With the chosen ordering convention the charge $q$
measures precisely the extra coupling to the U(1) part of the connection.
The quantum covariant momenta are hermitan conjugate to each other when using the
covariant version of the inner product in (\ref{Hmetric}), namely
\begin{equation}
\langle \chi|\phi \rangle = \int dx d\bar x\, g\, d \psi d\bar \psi\;  {\rm e}^{\bar \psi \psi}\;
\overline{\chi(x,\bar x, \psi)}\; \phi(x,\bar x, \psi)
\end{equation}
where $g=\det g_{\mu\bar \nu}$.
Note that with this inner product the hermiticity property of the momentum reads:
$p_\mu^\dagger=  \bar p_{\bar \mu} + i g_{\lambda\bar \lambda}g^{\nu\bar \nu}
\Gamma^{\bar\lambda}_{\bar \mu \bar \nu} \psi^\lambda \bar \psi_\nu$.

At this point one is ready to recognize the quantum version of the supersymmetric charges
\begin{equation}
\begin{split}
&
Q_{cl}= \psi^\mu\pi_\mu   \quad  \quad \ \to \quad
Q = \psi^\mu g^{1/2}\pi_\mu g^{-1/2}=
\psi^\mu g^{1/2} \left (p_\mu -i q \Gamma_\mu \right ) g^{-1/2}\\[1mm]
& \bar Q_{cl}= \bar\psi_\mu g^{\mu\bar\nu}\bar \pi_{\bar\nu}
\quad    \to \quad
\bar Q= \bar\psi_\mu g^{\mu\bar\nu} g^{1/2} \bar \pi_{\bar\nu} g^{-1/2}
= \bar\psi_\mu g^{\mu\bar\nu} g^{1/2}
\left (\bar p_{\bar\nu} +i q \Bar \Gamma_{\bar \nu} \right )
 g^{-1/2} \;.
\end{split}
\end{equation}
The powers of $g$ are required to obtain the correct hermiticity properties.
Again, the Christoffel connection drops out form the supercharge $Q$, as in the classical case.
As the $\psi$'s can be represented by the  coordinate basis of the $(1,0)$-forms, $\psi^\mu=dx^\mu$,
while their momenta as formal derivatives thereof, $\bar\psi_\mu=\frac{\de}{\de(dx^\mu)}$, we recognize that
 the supercharge $Q$ is represented by the Dolbeault operator twisted by the $U(1)$ connection
\begin{equation}\label{quantum Q}
iQ=i\psi^\mu\pi_\mu=\de_q\equiv\de+ q\Gamma\;,
\end{equation}
where $\Gamma= \Gamma_\mu dx^\mu=\Gamma^\nu_{\nu\mu}dx^\mu$
is the $U(1)$ connection form, and obeys $\de_q^2=0$.
Conversely, the charge $\bar Q$ is given by a twisted divergence
\begin{equation}\label{quantum Q bar}
i\bar Q=i\bar\psi_\mu\,g^{\mu\bar\nu}\bar\pi_{\bar\nu}
=-\de^\dagger_q\equiv\frac{\de}{\de(dx^\mu)}g^{\mu\bar\nu}(\bar \de_{\bar\nu} -q \bar \Gamma_{\bar\nu})\;.
\end{equation}
Thus, the quantum supercharges are conjugates under the adjoint operation, $Q^\dagger=\bar Q$, and define a self adjoint hamiltonian
\begin{equation}\label{quantum H}
\begin{split}
H_q &=\{Q,\bar Q\} =  \de_q\de^\dagger_q+ \de^\dagger_q\de_q
\\[1mm]&=
\frac12g^{\mu\bar\nu}g^{1/2}\big(\pi_\mu\bar\pi_{\bar\nu}+\bar\pi_{\bar\nu}\pi_\mu\big)g^{-1/2}+\frac12(1-4q)R^\mu_\nu\,\psi^\nu\bar\psi_\mu
+q\,R\\[1mm]
&= -\frac12\nabla^2_q+\frac12(1-4q)R^\mu_\nu\,dx^\nu\frac{\de}{\de (dx^\mu)}+q\,R\;,
\end{split}
\end{equation}
where the laplacian dressed with the $U(1)$ charge $q$ reads
$$
\nabla^2_q\equiv g^{\mu\bar\nu}[(\nabla_\mu+q\Gamma_\mu) (\bar \nabla_{\bar\nu}-q\bar\Gamma_{\bar\nu})
+(\bar \nabla_{\bar\nu}-q\bar\Gamma_{\bar\nu})(\nabla_\mu+q\Gamma_\mu)]\;.
$$
Let us notice that for the choice $q=\frac14$ the coupling to the Ricci tensor disappears, and the hamiltonian reduces to the square 
of the Dirac operator, as outlined in appendix \ref{app:Dirac},
$H_{1/4}=\frac12 g^{1/2}\pi^2_{\mathrm{sym}}g^{-1/2}+\frac14 R$.

By means of the differential operators just introduced the Maxwell-like equations for the $(p+1,0)$ curvature form read as
\begin{equation}
\de_q F_{(p+1,0)}=0\;,\quad \de^\dagger_qF_{(p+1,0)}=0\;.
\end{equation}
As in flat space, the first one can be integrated by introducing a $(p,0)$-form gauge field: $F_{(p+1,0)}=\de_q A_{(p,0)}$, 
defined up to gauge transformations $\delta A_{(p,0)}=\de_q\lambda_{(p-1,0)}$. The field equations then read
$\de^\dagger_q\de_q A_{(p,0)}=0$, and are a natural generalization of Maxwell's equations.
If desired, one may extract the laplacian $\nabla^2_q$ and cast them in the alternative form
\begin{equation}\label{A equation}
\Big(-\tfrac12\nabla^2_q+qR\Big)A_{(p,0)}+\frac{p}{2}(1-4q)\, \mathbf{Ric}\cdot A_{(p,0)}-\de_q\de^\dagger_qA_{(p,0)}=0\;,
\end{equation}
with $\mathbf{Ric}\cdot A_{(p,0)} \equiv R^\lambda_{\mu_1}A_{\lambda\mu_2...\mu_p}dx^{\mu_1}\wedge...\wedge dx^{\mu_p}$.

At the present stage, it is useful to study the transition amplitude associated to the quantum hamiltonian \eqref{quantum H}, as it will be of 
primary importance in the set up of the correct path integral that is needed in subsequent  applications, such as the evaluation of the effective action
of the $(p,0)$-form gauge fields.
One can evaluate the matrix element of the euclidean evolution operator between position eigenstates and coherent states for fermionic variables, 
$\Braket{x\bar\eta}{y\xi}{e^{-\beta H_q}}$, as a perturbative expansion in $\beta$. As usual, the calculation can be performed either 
by operatorial or functional methods. The operatorial computation, that makes use of the fundamental (anti)-commutation relations, 
is more involved, but it gives a completely non-ambiguous result for the transition amplitude  and can be used as a bench mark for setting up 
the path integral. Following the same computational method illustrated in \cite{Peeters:1993vu,Bastianelli:2011cc} for generic curved spaces, 
and in \cite{Bastianelli:2010ir} for models on K\"ahler manifolds,
we find the transition amplitude associated to the hamiltonian \eqref{quantum H}, up to first order in $\beta$. 
We restrict ourselves to the computation at coincident points, which is enough for our purposes, and find
\begin{equation}\label{TA operator}
\Braket{x\bar\eta}{x\xi}{e^{-\beta H_q}}=(2\pi\beta)^{-d}e^{\bar\eta\cdot\xi}\Big\{1+\beta\Big[(q-\tfrac13)R
+\frac12(4q-1)R_{\mu\bar\nu}\,\xi^\mu\bar\eta^{\bar\nu}\Big]+{\cal O}(\beta^2) \Big\}\;.
\end{equation}
Let us now turn to the functional computation. The classical hamiltonian corresponding to \eqref{quantum H} is given by
eq. \eqref{3.2}, and produces the
configuration space action \eqref{ungauged}. If we perform the path integral quantization by using the action \eqref{ungauged},
and regulate it to maintain covariance,
 it is natural to expect that a well defined quantum charge 
 for the U(1) subgroup of the holonomy group
 will be reproduced. In order to keep room for an arbitrary charge $q$, we dress the path integral action with a ``gauge field'' 
 counterterm proportional to an extra coupling $q_1$
\begin{equation}\label{ungauged with extra q}
S=\int dt\Big[g_{\mu\bar\nu}\dot x^\mu\dot{\bar{x}}^{\bar\nu}+i\bar\psi_\mu D_t\psi^\mu+iq_1\dot x^\mu\Gamma_\mu
-iq_1\dot{\bar x}^{\bar\mu}\bar\Gamma_{\bar\mu}+2q_1R^\mu_\nu\,\psi^\nu\bar\psi_\mu\Big]\;,
\end{equation}
whose structure is dictated by reality of the action and supersymmetry at the classical level. At this juncture we can evaluate the transition 
amplitude $\Braket{x\bar\eta}{x\xi}{e^{-\beta H_q}}$ by means of a functional integral suitably regulated (we use TS regularization, which 
generically requires only
covariant counterterms on K\"ahler manifolds, but MR and DR could be used as well, see \cite{Bastianelli:2006rx,Bastianelli:2011cc})
 giving at order $\beta$
\begin{equation}\label{TA path integral}
\Braket{x\bar\eta}{x\xi}{e^{-\beta H_q}}=(2\pi\beta)^{-d}e^{\bar\eta\cdot\xi}\Big\{1+\beta\Big[(q_1-\tfrac{1}{12})R+2q_1\,
R_{\mu\bar\nu}\,\xi^\mu\bar\eta^{\bar\nu}\Big] +{\cal O}(\beta^2) \Big\}\;.
\end{equation}
By comparing the two results \eqref{TA operator} and \eqref{TA path integral} we can exploit the relation among the true quantum charge $q$ 
and the counterterm one $q_1$. The path integral with action \eqref{ungauged} without extra charges ($q_1=0$) reproduces $q=\tfrac14$,
and more generally it follows that $q_1=q-\tfrac14$. This allows to keep control on the precise $U(1)$ couplings of the model 
in all the subsequent applications.


We end up this section with a  review of the calculation of the Witten index identified by the present supersymmetric sigma model,
as it will enter subsequent analyses. It yields the topological index of the (twisted) Dirac operator on K\"ahler manifolds.
The basics of this calculation were originally presented in \cite{AlvarezGaume:1983at, Friedan:1983xr},
and analyzed more recently in \cite{Ivanov:2010ki, Smilga:2011wi}.
The connection between index theorems and
supersymmetric quantum mechanics makes use of the concept of the Witten index, defined
as ${\rm Tr}\: (-1)^F$, where $F$ is the fermion number and the trace is over the quantum mechanical
Hilbert space.
Standard reasonings show that the Witten index
counts the number of bosonic zero energy states minus the number of fermionic zero energy states
\cite{Witten:1982df}. It is a topological invariant that computes the index of the differential operator
representing the hermitian supercharge $Q+\bar Q$. For the value $q=\frac14$, that we analyze first, it realizes
the Dirac operator $\Dslash \sim \partial_{\frac14} +\partial^\dagger_{\frac14}$, see appendix \ref{app:Dirac}.
In the Hilbert space of the particle system, bosonic states are given by $(p,0)$-forms with even $p$,
and fermionic states by forms with odd $p$. They correspond to positive chirality
and negative chirality spinors, respectively.
Thus for our quantum mechanical model the Witten index reduces to the Dirac index.
Being a topological invariant it can be regulated as  ${\rm Tr}\: (-1)^F {\rm e}^{-\beta H}$,
where $H$ is the hamiltonian, and computed for small $\beta$ using its path integral
representation
\begin{equation} \label{index}
{\rm ind} (\Dslash\,) = {\rm Tr}\: (-1)^F {\rm e}^{-\beta H}=
\int_P Dx D\psi\;  {\rm e}^{-S}
\end{equation}
where the subscript $P$ indicates periodic boundary conditions for bosonic and fermionic fields,
and $S$ is the Wick rotated version of the action in (\ref{ungauged}), namely
\begin{equation} \label{ungauged wick rotated}
S = \int_0^\beta d\tau\,\Big[g_{\mu\bar\nu}\dot x^\mu \dot{\bar x}^{\bar\nu}
+\bar\psi_\mu D_{\tau}\psi^\mu\Big] \;.
\end{equation}
The pure Dirac case is given by $q=\frac14$, and thus $q_1=0$, so that we
disregard the counterterms inserted in (\ref{ungauged with extra q}).
To calculate (\ref{index}) one expands all periodic fields in Fourier series with frequencies
$\frac{2\pi n}{\beta}$. For small $\beta$ the zero modes dominate, and one only needs to
take care of the semiclassical corrections due to a bosonic determinant.
It is useful to use Riemann normal coordinates adapted to the K\"ahler structure, scale suitably
the fermionic zero mode by $\beta^{-\frac12}$,  and obtain
\begin{equation} \label{determinants}
{\rm ind} ({\Dslash\,})=\int \frac{d^{d}x_0 d^{d}\bar x_0
d^d\psi_0 d^d\bar \psi_0 }{(2\pi)^d}\;
\left [\frac{{\rm Det}' (-\partial_\tau^2 + \cal{R}\partial_\tau)}{{\rm Det}' (-\partial_\tau^2 )}
\right ]^{-1}
\end{equation}
where ${\rm Det}'$ indicates a functional determinant on the space of periodic
fields orthogonal to the zero modes, the subscript 0 indicates zero modes,
and ${\cal R}=R^\mu{}_{\nu\bar \lambda\sigma} \bar \psi_0^{\bar\lambda} \psi^\sigma_0$
describes a matrix valued two-form evaluated at the point $(x_0,\bar x_0)$.
Now one can compute the functional determinant and express it in terms of a standard
$d \times d$ determinant of a matrix given by a function of ${\cal R}$
\begin{equation}
\frac{{\rm Det}' (-\partial_\tau^2 + \cal{R}\partial_\tau)}{{\rm Det}' (-\partial_\tau^2 )}
= {\rm det}  \left (\frac{\sinh {\cal R}/2}{{\cal R}/2} \right ) \;.
\end{equation}
Berezin integration over the Grassmann variables extracts from the expansion of the determinant
the contribution of the top $2d$-form only.
Thus one can reabsorb  the measure factor\footnote{This is $(2 \pi i)^d$ when taking into account 
the choice of a suitable orientation and the factors of $i$ 
present in the measure  (\ref{def-measure}).}
 into the determinant and present the final answer as
\begin{equation} \label{a roof genus}
{\rm ind} (\Dslash\,)=\int_M {\rm det}  \left (\frac{{\cal R}/4\pi i}{\sinh {\cal R}/4\pi i}\right )
\end{equation}
where now ${\cal R}=R^\mu{}_{\nu\bar \lambda\sigma} d\bar x^{\bar \lambda} d x^\sigma$.

As just mentioned, for a given K\"ahler  manifold $M$ only the top form coming from the expansion of the determinant
contributes. Since the determinant of an even function of ${\cal R}$ has an expansion in terms of  ${\cal R}^2$,
the index is nonvanishing only for manifolds of even complex dimensions.
The first example is for $d=2$, where the above formula gives
\begin{equation} \label{example}
{\rm ind} (\Dslash\,)= \frac{1}{96\pi^2} \int_M {\rm tr}\  {\cal R}^2=
\frac{1}{96\pi^2} \int_M
d{\bar x}^1d{\bar x}^2dx^1dx^2 g \left ( 
R_{\mu\bar \nu \lambda \bar \sigma} R^{\mu\bar \nu \lambda \bar \sigma}
- R_{\mu\bar \nu} R^{\mu\bar \nu} \right ).
\end{equation}

In general we are interested in keeping an arbitrary U(1) charge in the twisted Dolbeault operators
$\partial_q$ and $\partial^\dagger_q$. For $q\neq \frac14$ this identifies a sort of twisted Dirac operator,
which we denote by $\Dslash_{q_1}$ (so that $\Dslash_{\,0} = \Dslash$\,).
To compute its index we have to dress the previous computation by considering the counterterms  proportional to $q_1$
in (\ref{ungauged with extra q}), which under Wick rotation produce
\begin{equation}
\Delta S = \int_0^\beta d\tau
\Big[ q_1\dot x^\mu\Gamma_\mu-q_1\dot{\bar x}^{\bar\mu}\bar\Gamma_{\bar\mu}-2q_1R^\mu_\nu\,\psi^\nu\bar\psi_\mu\Big]\;.
\end{equation}
Suitably rescaling the quantum fields as described above, one recognizes that only the last term may contribute
through its leading expansion around the zero modes. This appears in the exponential of the path integral as
\begin{equation}
{\rm e}^{2 q_1 R_{\mu\bar\nu}(x_0,\bar x_0)\psi_0^\mu \bar \psi_0^{\bar \nu}}
\end{equation}
which must be inserted inside the integral of eq. (\ref{determinants}).
The final formula for the twisted Dirac operator is then
\begin{equation}
{\rm ind} (\Dslash_{q_1})=\int_M  \exp{\left (\frac{ q_1 \cal{F}}{\pi i}\right)} \, {\rm det}  \left (\frac{{\cal R}/4\pi i}{\sinh {\cal R}/4\pi i}\right )
\end{equation}
with ${\cal{F}}=R_{\mu\bar\nu} d x^\mu d\bar x^{\bar \nu} $. In $d=2$ it produces the following extra contribution
\begin{equation} 
 \frac{q_1^2}{2\pi^2}  \int_M
d{\bar x}^1d{\bar x}^2dx^1dx^2 g \left ( R^2 - R_{\mu\bar \nu} R^{\mu\bar \nu} \right ).
\end{equation}
that added to (\ref{example}) gives the index of the twisted Dirac operator ${\rm ind} (\Dslash_{q_1})$.

\section{Effective action of quantized $\bf (p,0)$-forms}
We are now ready to come to the main part of the paper, discuss the quantization
of $(p,0)$-forms and compute the corresponding effective actions using worldline methods.

To stat with, we aim at obtaining a useful worldline representation of the one-loop effective action
in an arbitrary K\"ahler background. The effective action may be depicted by the sum of all
Feynman diagrams of the form shown in figure 1, where a quantum  $(p,0)$-form gauge field circulates in the loop
and external lines represent the curved background.
\begin{figure}[h]
  \centering
  \includegraphics[width=1.8in]{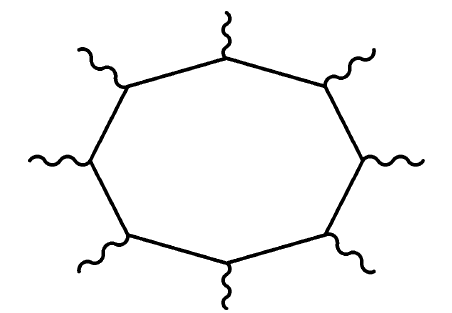}
  \caption[Fig1]%
  {Feynman diagram for the one-loop effective action. A quantum  $(p,0)$-form circulates in the loop
and external lines represent the curved background.}
\end{figure}

In first quantization the physical degrees of freedom are carried by a quantum spinning particle
that circulates in the loop. As we shall see,
the worldline representation allows to study various quantum properties and derive precise duality relations.
Generically one is not able to compute the effective action exactly, but
one may try to compute it in some perturbative expansion.
Here, we compute the first few heat kernel coefficients that are identified by a short proper time expansion.

As discussed, to obtain the  Maxwell equation for a  $(p,0)$-form gauge field from the particle system we need to gauge the
whole  $U(1)$ supersymmetry algebra carried by the ungauged model of eq. (\ref{3.3}).
The suitable covariantization of the charges has been
described in the previous section, see eqs.  (\ref{3.1}) and (\ref{3.2}).
The action with local symmetries is obtained by coupling worldline gauge fields to the charges
and adding a Chern-Simons coupling $s$.
Thus one obtains an action of the same form as (\ref{spinning particle}), but with covariantized charges.
To recover the euclidean action in configuration space we first eliminate momenta by means of their equations of motion, 
and then perform a Wick rotation, obtaining
\begin{equation} \label{euclidean action}
\begin{split}
S[X,G] &= \int_0^1 d\tau\,\Big[e^{-1}g_{\mu\bar\nu}\big(\dot x^\mu-\bar\chi\psi^\mu\big)\big(\dot{\bar x}^{\bar\nu}
-\chi\bar\psi^{\bar\nu}\big)+\bar\psi_\mu\big[D_{\tau}+ia\big]\psi^\mu+isa\Big]\\
&+q_1\int_0^1d\tau\Big[\dot x^\mu\Gamma_\mu-\dot{\bar x}^{\bar\mu}\bar\Gamma_{\bar\mu}-2e\,R^\mu_\nu\,\psi^\nu\bar\psi_\mu\Big]\;,
\end{split}
\end{equation}
where we recall that a counterterm proportional to $q_1 \equiv q-\frac14$ is needed in order to reproduce
a quantum coupling $q$ to the $U(1)$ part of the connection. We denote the basic dynamical variables by
$X=(x^\mu,\bar x^{\bar \mu}, \psi^\mu, \bar\psi_\mu)$ and $G=(e,\chi,\bar\chi,a)$. Of course
$\bar\psi^{\bar\nu}=g^{\mu\bar\nu}\bar\psi_\mu$, while the covariant time derivative is given by
$D_\tau\psi^\mu=\dot\psi^\mu+\dot x^\nu\Gamma^\mu_{\nu\lambda}\psi^\lambda$.
Note that along with the Wick rotation $t\to -i\tau$,
we have rotated also the gauge field $a \to ia$ to keep the U(1) gauge group compact.

Quantization of this spinning particle model on a circle parametrized by $\tau\in[0,1]$ gives the partition function for the
$(p,0)$-form gauge field coupled to the metric of the curved K\"ahler space
\begin{equation}\label{partition function defined}
Z[g]\propto\int\frac{\D X\D G}{\text{Vol(Gauge)}}\, {\rm e}^{-S[X,G]}
\end{equation}
and visually corresponds to figure 1.
A point worth stressing again is that we regulate the path integral and related
functional determinants so that they correspond to a graded-symmetric operatorial
ordering of the current $J$,
namely $J= \frac12 (\psi^\mu\bar \psi_\mu -\bar \psi_\mu\psi^\mu)= \psi^\mu\bar \psi_\mu -\frac{d}{2}$,
an ordering that is responsible, for example, to the standard fermionic zero point energy.
Then the projection onto the physical field strenght $F_{(p+1,0)}$ is obtained by using the
Chern-Simons coupling $s \equiv p+1-\frac{d}{2}$ (so that $J-s=\psi\bar\psi -(p+1)$ as an operator).

Using the standard Fadeev-Popov procedure to get rid of gauge redundancy, we fix gauge fields to the constant values
$\tilde G=(\beta,0,0,\phi)$, and are left with modular integrations over $\beta$ and $\phi$, with the following one-loop measure 
that was carefully studied in \cite{Bastianelli:2005vk}
\begin{equation}\label{partition function}
Z[g]\propto\int_0^\infty\frac{d\beta}{\beta}\int_0^{2\pi}\frac{d\phi}{2\pi}\left(2\cos\frac\phi2\right)^{-2}\int_{\text{P}}\D x\D\bar x\int_{\text{A}}
D\bar\psi D\psi\, {\rm e}^{-S[X,\tilde G]}
\end{equation}
with $S[X,\tilde G]$
denoting the gauge fixed action, i.e. eq. (\ref{euclidean action})
evaluated at $G=\tilde G$. The subscript P and A denote periodic and antiperiodic boundary conditions, respectively.
The integral over $\beta$ is the usual proper time integral with the well known one-loop measure, while the factor $\left(2\cos\frac\phi2\right)^{-2}$ 
is the Fadeev-Popov determinant of the bosonic superghosts associated to $\chi$ and $\bar\chi$.
We denote with $\D x$ the general coordinate invariant measure, i.e.
$\D x \D \bar x \sim\prod_{\tau=0}^1d^dx(\tau)\, d^d\bar x(\tau)\, g(x(\tau), \bar x(\tau))$, with $g=\det g_{\mu\bar\nu}$,
while $D\psi \sim\prod_{\tau=0}^1d^d\psi(\tau)$
is the simple translational invariant measure.\footnote{Note that, since $\psi$'s are spacetime vectors, while $\bar\psi$'s are covectors, 
one has $\D\bar\psi\D\psi=D\bar\psi D\psi$.} This formula gives the worldline representation of the effective action
of the $(p,0)$-form gauge field.

For computational purposes, it is useful to manipulate it a bit further. The path integral over loops,
i.e. over coordinate fields with periodic boundary conditions, can be done in several ways \cite{Bastianelli:2003bg}.
Here we choose to fall back on quantum fields with Dirichlet boundary conditions. Thus,
we pick an arbitrary $x_0$ as a base-point for our loops. 
The path integral then factorizes as $\int_{\text{P}}\D x\D\bar x=\int d^dx_0d^d\bar x_0g(x_0)\int_{x(0)=x(1)=x_0}\!\!\!\!\D x\D\bar x$. 
It is possible then to perform background-quantum fluctuations splitting as $x^\mu(\tau)=x_0^\mu+q^\mu(\tau)$, with $q^\mu(0)=q^\mu(1)=0$. 
Clearly the $x$ path integral becomes $\int_{\text{D}}\D q\D\bar q$, where D stands for Dirichlet boundary conditions, i.e. fields
 are taken to vanish at boundaries. The next step is that of getting rid of the field dependent measure $\D q\D\bar q$.
Following the trick of \cite{Bastianelli:1991be, Bastianelli:1992ct} we
exponentiate the $g$ factors with a path integral over fermionic complex ghosts $b^\mu$ and $\bar c^{\bar\nu}$: 
$\D q\D\bar q=DqD\bar q\int DbD\bar c\; {\rm e}^{-S_{\text{gh}}}$. At this stage the gauge fixed action
$S_{\text{gf}}\equiv S[X,\tilde G]$ plus
the ghost action for the path integral measure $S_{\text{gh}}$
take the following form\footnote{We rescaled fermions by
$\psi\to \frac{1}{\sqrt\beta}\psi$ in order to extract a common $\beta$ as loop counting parameter.}
\begin{equation}
\begin{split}
S_{\text{gf}}+S_{\text{gh}} &= \frac1\beta\int_0^1d\tau\,\Big[g_{\mu\bar\nu}\big(\dot q^\mu\dot{\bar q}^{\bar\nu}
+b^\mu\bar c^{\bar\nu}\big)+\bar\psi_\mu(D_\tau+i\phi)\psi^\mu\\
&+\beta q_1\big(\dot q^\mu\Gamma_\mu-\dot{\bar q}^{\bar\mu}\bar\Gamma_{\bar\mu}-2R^\mu_\nu\,\psi^\nu\bar\psi_\mu\big)
\Big]+is\phi \;.
\end{split}
\end{equation}

In order to perform perturbative calculations we expand all background fields around the fixed point $x_0$. 
The action written above splits into a quadratic part $S_2$ giving propagators, as usual, and an interaction part. 
We denote as $\media{\,\bullet\,}$ the quantum average weighted with the free path integral:
 $\media{\,\bullet\,}=\frac{1}{\int {\rm e}^{-S_2}}\int\,\bullet\ {\rm e}^{-S_2}$. The partition function \eqref{partition function} now reads
\begin{equation}\label{ea}
Z\propto\int_0^\infty\frac{d\beta}{\beta}\int_0^{2\pi}\frac{d\phi}{2\pi}\left(2\cos\frac\phi2\right)^{d-2}
{\rm e}^{-is\phi}\int\frac{d^dx_0d^d\bar x_0}{(2\pi\beta)^d}g(x_0)\media{{\rm e}^{-S_{\text{int}}}}\;,
\end{equation}
where $(2\cos\frac\phi2)^d(2\pi\beta)^{-d}$ is the usual free path integral normalization, and the interaction part is
\begin{equation}
\begin{split}
S_{\text{int}} &= \frac1\beta\int_0^1d\tau\,\Big[\big(g_{\mu\bar\nu}(x_0+q)-g_{\mu\bar\nu}(x_0)\big)\big(\dot q^\mu\dot{\bar q}^{\bar\nu}
+b^\mu\bar c^{\bar\nu}\big)+\dot q^\nu\Gamma^\mu_{\nu\lambda}(x_0+q)\bar\psi_\mu\psi^\lambda\\
&+\beta q_1\big(\dot q^\mu\Gamma_\mu(x_0+q)-\dot{\bar q}^{\bar\mu}\bar\Gamma_{\bar\mu}(x_0+q)-2R^\mu_\nu(x_0+q)\,
\psi^\nu\bar\psi_\mu\big)\Big].
\end{split}
\end{equation}
For our computation we can choose any coordinate system so, in order to be able to reconstruct covariance,
and at the same time to maintain holomorphic  coordinates, we use K\"ahler normal coordinates
(see \cite{Higashijima:2000wz}, for example) centered at $x_0$.
Denoting with $S_n$ the part of $S_{\text{int}}$ containing $n$-fields vertices (or less, but producing a result of the same order
in $\beta$), it results that, in K\"ahler normal coordinates, 
the only terms giving non vanishing contribution up to order $\beta^2$ are the following ones
\begin{equation}\label{S_4 and S_6}
\begin{split}
S_4 &= \frac1\beta\int_0^1 d\tau\,\Big[R_{\mu\bar\nu\lambda\bar\sigma}\,q^\lambda\bar q^{\bar\sigma}\big(\dot q^\mu\dot{\bar q}^{\bar\nu}
+b^\mu\bar c^{\bar\nu}\big)+R^\lambda{}_{\sigma\bar\nu\mu}\,\dot q^\mu\bar q^{\bar\nu}\bar\psi_\lambda\psi^\sigma\Big]\\
&+q_1\int_0^1d\tau\Big[R_{\mu\bar\nu}\big(q^\mu\dot{\bar q}^{\bar\nu}-\dot q^\mu\bar q^{\bar\nu}\big)-2\,R^\mu_\nu\,\psi^\nu\bar\psi_\mu\Big]\;,\\
S_6 &= \frac1\beta\int_0^1 d\tau\,\Big[\frac14\big[\nabla_{(\bar\sigma}\nabla_\lambda R_{\mu\bar\nu\rho\bar\kappa)}
+3R^{\bar\tau}{}_{(\bar\nu\lambda\bar\kappa}R_{\mu\bar\sigma\rho)\bar\tau}\big]q^\lambda\bar q^{\bar\sigma}q^\rho\bar q^{\bar\kappa}
\big(\dot q^\mu\dot{\bar q}^{\bar\nu}+b^\mu\bar c^{\bar\nu}\big)\\
&-\frac12\big[\nabla_\rho\nabla_{\bar\sigma}R^\lambda{}_{\mu\bar\lambda\nu}+R^{\bar\tau}{}_{\bar\lambda\rho\bar\sigma}
R^\lambda{}_{\mu\bar\tau\nu}\big]q^\rho\bar q^{\bar\sigma}\bar q^{\bar\lambda}\dot q^\mu\psi^\nu\bar\psi_\lambda\Big]\\
&-q_1\int_0^1d\tau\Big[\frac12\big[\nabla_\rho\nabla_{\bar\sigma}R_{\mu\bar\lambda}+R^{\bar\tau}{}_{\bar\lambda\rho\bar\sigma}
R_{\mu\bar\tau}\big]q^\rho\bar q^{\bar\sigma}\bar q^{\bar\lambda}\dot q^\mu
\\ & -\frac12\big[\nabla_{\bar\rho}\nabla_{\sigma}R_{\bar\mu\lambda}+R^{\tau}{}_{\lambda\bar\rho\sigma}
R_{\bar\mu\tau}\big]\bar q^{\bar\rho} q^\sigma q^{\lambda}\dot{\bar q}^{\bar\mu}
 +2\nabla_\lambda\nabla_{\bar\sigma}R^\mu_\nu\,q^\lambda\bar q^{\bar\sigma}\psi^\nu\bar\psi_\mu\Big]\;,
\end{split}
\end{equation}
where all tensors are calculated at $x_0$ and round brackets denote weighted symmetrization, separately among 
holomorphic and anti-holomorphic indices, i.e. $A_{(\mu_1...\mu_n\bar\nu_1...\bar\nu_m)}\equiv A_{(\mu_1...\mu_n)(\nu_1...\nu_m)}$.
From the quadratic action 
$S_2=\frac1\beta\int[g_{\mu\bar\nu}(x_0)(\dot q^\mu\dot{\bar q}^{\bar\nu}+b^\mu\bar c^{\bar\nu})+\bar\psi_\mu(\de_\tau+i\phi)\psi^\mu]$ 
one extracts the following two point functions
\begin{equation}\label{2 point functions}
\begin{split}
\media{q^{\mu}(\tau)\bar q^{\bar\nu}(\sigma)} &= -\beta g^{\mu\bar\nu}(x_0)\Delta(\tau,\sigma)\;,
\quad\media{b^\mu(\tau)\bar c^{\bar\nu}(\sigma)}=-\beta g^{\mu\bar\nu}(x_0)\delta(\tau,\sigma) \\[1mm]
\media{\psi^\mu(\tau)\bar\psi_\nu(\sigma)} &= \beta \delta^\mu_\nu\Delta_f(\tau-\sigma,\phi)
\end{split}
\end{equation}
where the propagators in the continuum limit read
\begin{equation}\label{propagators}
\begin{split}
\Delta(\tau,\sigma) &= \sigma(\tau-1)\,\theta(\tau-\sigma)+\tau(\sigma-1)\,\theta(\sigma-\tau)\;,\\
\Delta_f(x,\phi) &= \frac{{\rm e}^{-i\phi x}}{2\cos\frac\phi2}\big[{\rm e}^{i\frac\phi2}\theta(x)-{\rm e}^{-i\frac\phi2}\theta(-x)\big]
\end{split}
\end{equation}
with $\theta(x)$ the step function and  $\delta(\tau,\sigma)$ the Dirac delta
acting on functions vanishing at the boundaries. We note that in performing perturbative calculations one encounters products 
and derivatives of such distributions, that are ill defined. To resolve this ambiguity we use Time Slicing (TS) regularization
\cite{DeBoer:1995hv, deBoer:1995cb, Bastianelli:2006rx},
that gives well-known prescriptions on how to handle such products of distributions
and necessitates no counterterms (the standard TS counterterm vanish on K\"ahler manifolds).
The rules are as follows: when computing the various Feynmann diagrams all delta functions
should be implemented with the prescription of considering $\theta(0)=\frac12$ for the step function,
while the ghost system guarantees that no products of delta functions can ever arise.

Looking at \eqref{2 point functions} we immediately see that each piece $S_n$ of $S_{\text{int}}$ gives a contribution 
of order $\beta^{n/2-1}$. Therefore, our quantum average can be written explicitly as
\begin{equation}
\media{{\rm e}^{-S_{\text{int}}}}=1-\media{S_4}-\media{S_6}+\frac12\media{S_4^2}+{\cal O}(\beta^3)\;.
\end{equation}
Using the expressions given in \eqref{S_4 and S_6} and TS prescriptions in calculating Feynman diagrams, one finally obtains
\begin{equation}\label{master formula U(1)}
\begin{split}
\media{{\rm e}^{-S_{\text{int}}}} &= 1+\beta\Big(iq_1\tan\frac\phi2-\frac{1}{12}\Big)\,R+\beta^2\,\Big\{\Big(\frac{1}{180}
-\frac{1}{96}\cos^{-2}\frac\phi2\Big)\,R_{\mu\bar\nu\lambda\bar\sigma}R^{\mu\bar\nu\lambda\bar\sigma}\\
&+\Big[-\frac{19}{1440}-\frac{q_1^2}{6}+\Big(\frac{1}{96}+\frac{q_1^2}{2}\Big)\cos^{-2}\frac\phi2+\frac{iq_1}{12}\tan\frac\phi2\Big]\,
R_{\mu\bar\nu}R^{\mu\bar\nu}\\
&+\Big(\frac{1}{288}+\frac{q_1^2}{2}-\frac{q_1^2}{2}\cos^{-2}\frac\phi2-\frac{iq_1}{12}\tan\frac\phi2\Big)\,R^2
\\&
+\Big(-\frac{1}{240}+\frac{iq_1}{12}\tan\frac\phi2\Big)\, \nabla^2R\Big\}\;,
\end{split}
\end{equation}
where $\nabla^2R=2g^{\mu\bar\nu}\de_\mu\de_{\bar\nu}R$.

Plugging this result into the partition function \eqref{partition function} one faces the task of
performing the $\phi$ integral, taking care of the possible pole arising at $\phi=\pi$. Switching to the Wilson loop variable $w=e^{i\phi}$ 
one has a contour integral on the unit circle surrounding the origin, with a possible pole on the integration path at $w=-1$. 
Its presence is related to topological mismatches, affecting duality relations, that we are going to investigate in the next section.
We need a prescription to deal with this additional pole, and the correct one turns out to be to slightly deform our path in a way
that excludes the pole, as shown in figure 2.  We call this regulated contour $\gamma^-$.
\begin{figure}[h]
  \centering
  \includegraphics[width=2.5in]{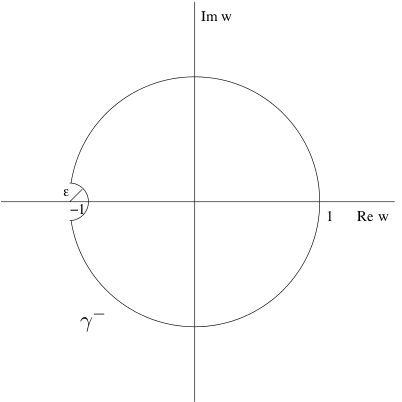}
  \caption[Fig2]%
  {The regulated contour $\gamma^-$ that excludes the pole at $w=-1$.}
\end{figure}
The correctness of this choice is confirmed by checking the result for a scalar field, that indeed comes out correctly
only by using the aforementioned prescription.

The additional pole at $w=-1$ shows up already at order $\beta^2$ for $d<4$, while for $d\geq4$ it appears at higher orders in $\beta$.
For this reason we present the results separately for $d\geq4$ and for lower dimensions, recalling that $(p,0)$-forms propagate only
for $0\leq p\leq d-2$. 
First of all, let us parametrize the structure of the first heat kernel coefficients as follows
\begin{equation}\label{Seeley coefficients parametrized}
Z\hspace{-1mm}\propto\hspace{-1mm}\int_0^\infty\hspace{-1mm}\frac{d\beta}{\beta}\hspace{-1mm}\int\frac{d^dx_0d^d\bar x_0}{(2\pi\beta)^d}
g(x_0)\Big\{v_1+v_2\beta\,R+\beta^2\Big[v_3\,
R_{\mu\bar\nu\lambda\bar\sigma}R^{\mu\bar\nu\lambda\bar\sigma}
+v_4\,R_{\mu\bar\nu}R^{\mu\bar\nu}+v_5\,R^2+v_6\,\nabla^2R\Big]\Big\}.
\end{equation}
Let us recall that the first coefficient $v_1$ in \eqref{Seeley coefficients parametrized} represents the number of physical degrees
of freedom, and will be zero when considering the contributions to the effective action of non-propagating fields.
We may now list the coefficients of a gauge $(p,0)$-form with charge $q$ 
in the format: $A^{(q)}_{p}\to(v_1;v_2;v_3;v_4;v_5;v_6)$,
where the $v_i$ are the coefficients appearing in eq. (\ref{Seeley coefficients parametrized}).

Let us start by giving the Seeley-DeWitt coefficients for a $(p,0)$-form in $d\geq4$
\begin{equation}\label{p form coefficients}
\begin{split}
A^{(q)}_{p}\to\ & \binom{d-2}{p}\times\left(1;\,\frac16-\frac{p}{2(d-2)}-q\frac{d-2-2p}{d-2};\,\frac{1}{180}-\frac{p(d-p-2)}{24(d-2)(d-3)};\right.\\[3mm]
&-\frac{1}{360}+\frac{p(3d-4p-5)}{24(d-2)(d-3)}+\frac{q}{6}\frac{p(6p-5d+9)}{(d-2)(d-3)}+
\frac{q^2}{6} \Big[\frac{12p(d-p-2)}{(d-2)(d-3)}-1\Big];\\[3mm]
&\frac{1}{72}+\frac{p(3p-2d+3)}{24(d-2)(d-3)}-\frac{q}{6}\Big[\frac{p(6p-5d+9)}{(d-2)(d-3)}+1\Big]
+\frac{q^2}{2} \Big[1-\frac{4p(d-p-2)}{(d-2)(d-3)}\Big];\\[3mm]
&\left.\frac{1}{60}-\frac{p}{24(d-2)}-\frac{q}{12}\frac{d-2-2p}{d-2}\right)\;.
\end{split}
\end{equation}
These are the coefficients for a gauge $(p,0)$-form coupled to the U(1) part of the connection via a charge $q$, 
obeying $\de^\dagger_q\de_q A_p=0$. They are invariant under the exchange $p\leftrightarrow (d-p-2)$ and $q\leftrightarrow \frac12-q$, 
as it is obvious if one rewrites them in terms of $q_1=q-\frac14$, the duality being $q_1\leftrightarrow -q_1$. 
This hints indeed towards a duality between $(p,0)$ and $(d-p-2,0)$-forms $A^{(q)}_p\leftrightarrow A_{d-p-2}^{(1/2-q)}$, that will be investigated 
in the next section.

We can immediately check that the result \eqref{p form coefficients} correctly reproduces the known coefficients for a scalar field: 
setting $p=0$ one gets
\begin{equation}\label{scalar}
A^{(q)}_0\to\left(1;\,\frac16-q;\,\frac{1}{180};\,-\frac{1}{360}-\frac{q^2}{6};\,\frac{1}{72}-\frac{q}{6}+\frac{q^2}{2};\,
\frac{1}{60}-\frac{q}{12}\right)\;,
\end{equation}
which coincide with the standard results\footnote{See appendix \ref{app:notations} to compare our conventions on curvatures with the 
standard riemannian ones.} once one turns off the charge $q$.

Let us examine a bit closer what happens in lower dimensions. In $d=3$ complex dimensions, only scalars and one-forms propagate. 
The formula \eqref{p form coefficients}, that is ill-defined for generic $p$ at $d=3$, has indeed a smooth limit for $p=0,1$ that reads
\begin{equation}\label{d=3 01 forms}
\begin{split}
&d=3\;,\quad p=0,1\\[2mm]
&A^{(q)}_{p}\to\left(1;\,\frac16-\frac p2+q(2p-1);\,\frac{1}{180}-\frac{p}{24};\,-\frac{1}{360}+\frac{p}{8}-\frac56 qp+\frac{q^2}{6}(12p-1);\right.\\[2mm]
&\left.\hspace{20mm}\frac{1}{72}-\frac{p}{12}+\frac q6 (5p-1)+\frac{q^2}{2}(1-4p);\,\frac{1}{60}-\frac{p}{24}+\frac{q}{12}(2p-1)\right)\;.
\end{split}
\end{equation}
In $d=3$ zero-forms are expected to be dual to one-forms, but \eqref{d=3 01 forms} is not invariant under the exchange 
$p\leftrightarrow 1-p$ and $q\leftrightarrow\frac12-q$. In fact, in $d=3$ the mismatches that are discussed in the next section appear already 
at order $\beta^2$. For $p>1$ the heat kernel coefficients are not zero in $d=3$, even though nothing propagates, and give just a 
topological contribution that will be exploited when addressing exact dualities.

A similar reasoning holds in $d=2$: now only scalars propagate, and equation \eqref{p form coefficients} has a smooth $d=2$ limit for 
$p=0$, yielding the known result \eqref{scalar}. 

Let us also discuss briefly the case of $d=1$, that is  somewhat degenerate.
The expansion of the generic wave function (\ref{wave function})
suggests as possible models those related to 
 $p=-1$ and $p=0$, as now one can write $\phi(x,\bar x, \psi)= F_0(x,\bar x) + F_1(x,\bar x)\psi$.
For each of them one of the susy constraint equations collapse  to an identity, and the remaining one 
corresponds to  $\partial_q F_0=0$ and $\partial_q^\dagger F_1=0$. In both cases one cannot legally introduce a gauge
potential $A_p$. Nevertheless the path integral computes their effective action, showing that for $p=-1$ (i.e. $F_0$)
the model is empty, while for $p=0$ (i.e. $F_1$) one obtains again the values of a scalar field as in eq. \eqref{scalar}. 

As another interesting application of our $U(1)$ spinning particle, we can choose not to gauge the $U(1)$ part of the first class algebra,
i.e. $J-s$. 
Then, we do not have a modular integration over $\phi$ any more, and the result for this new model is obtained for free by setting $\phi=0$ 
in \eqref{master formula U(1)}.
It corresponds to the quantum theory of the sum
of all $(p,0)$-forms $F_{p}$
with dynamics dictated by the Maxwell equations. We know that this system is equivalent, on K\"ahler manifolds, to a Dirac spinor; hence
its effective action must be proportional to the one-loop effective action of a Dirac field. In fact,
the path integral over the complex gravitino present in \eqref{partition function defined}
can at most change the overall normalization of the partition function if compared with the path integral over the real
gravitino needed for the Dirac field, recall eq. (\ref{spinning particle:Dirac}).
Indeed, one may check that fixing suitably the overall normalization, one recovers the heat kernel coefficients
of a Dirac spinor. In order to do so, we recall from previous sections that the sum $\de_q+\de^\dagger_q$ is equivalent to the Dirac operator 
only for $q=\frac14$, that is $q_1=0$. In terms of $q_1$, the heat kernel coefficients of the $U(1)$-ungauged model read
\begin{equation}\label{Dirac fermion coefficients}
\Psi^{(q_1)}\to \ 2^d\,\left(1;\,-\frac{1}{12};\,-\frac{7}{1440};\,-\frac{1}{360}+\frac{q_1^2}{3};\,\frac{1}{288};\,-\frac{1}{240}\right)\;,
\end{equation}
that indeed agree at $q_1=0$ with the standard results for a Dirac fermion, compare for example with \cite{dewitt, Bastianelli:2002qw}.

Finally, one might wish not to gauge the two supersymmetries at all, but gauge the $U(1)$ charge instead.
This produce the effective action of a single $(p,0)$-form $B_p$, now with dynamics
dictated by the hamiltonian $H_q$ only, namely a $(p,0)$-form without any gauge
invariance  but with dynamical equation $(\partial_q \partial_q^\dagger + \partial_q^\dagger
\partial_q) B_p=0$. To achieve this, we only need to drop from
(\ref{partition function}) the Faddeev-Popov determinat
$\left(2\cos\frac\phi2\right)^{-2}$ due to the gauge fixing of the gravitini,
fix the Chern-Simons coupling $s=p-\frac d2$, and obtain the following coefficients for the ``non gauge'' $(p,0)$-form 
$B_{p}$ with charge $q$
\begin{equation}\label{non gauge p form coefficients}
\begin{split}
B^{(q)}_{p}
\to\ & \binom{d}{p}\times\left(1;\,\frac16-\frac{p}{2d}+q\Big[\frac{2p}{d}-1\Big];\,\frac{1}{180}-\frac{p(d-p)}{24d(d-1)};\right.\\[3mm]
&-\frac{1}{360}+\frac{p(3d-4p+1)}{24d(d-1)}+\frac{q}{6}\frac{p(6p-5d-1)}{d(d-1)}+\frac{q^2}{6}\Big[\frac{12p(d-p)}{d(d-1)}-1\Big];\\[3mm]
&\frac{1}{72}+\frac{p(3p-2d-1)}{24d(d-1)}-\frac{q}{6}\Big[\frac{p(6p-5d-1)}{d(d-1)}+1\Big]+
\frac{q^2}{2}\Big[1-\frac{4p(d-p)}{d(d-1)}\Big];\\[3mm]
&\left.\frac{1}{60}-\frac{p}{24d}+\frac{q}{12}\Big[\frac{2p}{d}-1\Big]\right)\;.
\end{split}
\end{equation}
This formula is valid for $d>1$.
We notice that no additional pole arises at $w=-1$, and that the result \eqref{non gauge p form coefficients} is invariant under the 
simultaneous exchange 
$p\leftrightarrow d-p$ and $q\leftrightarrow \frac12-q$. This points towards a duality between $(p,0)$ and $(d-p,0)$ ``non gauge'' differential forms. In the special case of $d=1$ 
the Riemann and  Ricci tensor are not independent from the scalar curvature, so that it is enough to list the coefficients
for the $(p,0)$-forms in the format $(v_1;v_2;\bar v\equiv v_3+v_4+v_5;v_6)$ as 
$ R_{\mu\bar\nu\lambda\bar\sigma}R^{\mu\bar\nu\lambda\bar\sigma}=R_{\mu\bar\nu}R^{\mu\bar\nu}=R^2$.
The two possibilities are for $p=0,1$ and one gets
\begin{equation}
\begin{split}
B^{(q)}_{p}
\to\ & \left(1;\,\frac16-q + \frac{p}{2}(4q-1);\, \frac{1}{60} -\frac{q}{6}+\frac{q^2}{3};\,  
\frac{1}{60} -\frac{q}{12} + \frac{p}{24}(4q-1) \right)
\end{split}
\end{equation}
which signals a duality between $p=0$ and $p=1$.

\section{Dualities}
We now wish to  discuss in more depth the issue of duality, as emerged \lq\lq experimentally" form the results
of the last section. Here we prove exact relations between dual formulations.

It is useful to start with the classical particle action given in (\ref{euclidean action}),
which is characterized by the Chern-Simons coupling $s$ and the U(1) charge $q_1\equiv q-\frac14$.
One may begin by
noticing that the model with couplings $(-s, -q_1)$ is equivalent to the model with couplings $(s, q_1)$. In fact,
one obtains the latter from the former by a suitable transformation of the dynamical variables:
one needs to change the sign of the U(1) gauge field $ a\to -a$ (to bring the coupling $-s$ back to the value $+s$),
exchange $\psi \leftrightarrow \bar \psi$ (to bring the couplings of the gauge field $a$ to the fermions back to its  original form,
which contains a covariant derivative of the form $\partial_\tau + i a$),
and then exchange $x \leftrightarrow \bar x$ together with $\chi \leftrightarrow \bar \chi$
(to reinstate the correct overall $q_1$ coupling and  achieve at the same time full equivalence
with the $(s,q_1)$ model). Thus, one verifies that this change of variables relates the
model with couplings $(-s, -q_1)$ to the one with couplings $(s, q_1)$. At the quantum level
the equivalence between the two models corresponds  to a duality between different forms.

To discuss the latter is useful to switch to an operatorial picture and cast the effective
action (\ref{partition function}) as follows
\bea
Z_p(q)& \propto&
\int_0^\infty\frac{d\beta}{\beta}\int_0^{2\pi}\frac{d\phi}{2\pi}\left(2\cos\frac\phi2\right)^{-2}\int_{\text{P}}\D x\D\bar x\int_{\text{A}}
D\bar\psi D\psi\, {\rm e}^{-S[X,\tilde G]}
\label{5-path-int}
\\[2mm]
&=&
\int_0^\infty\frac{d\beta}{\beta}\int_0^{2\pi}\frac{d\phi}{2\pi}\left(2\cos\frac\phi2\right)^{-2}
{\rm Tr}\, [\, {\rm e}^{i\phi(J-s)} {\rm e}^{-\beta H_{q}}]
\label{5-oper-quant}
\\[2mm]
 &=& \int_0^\infty\frac{d\beta}{\beta}
\oint_{\gamma^-}  \frac{dw}{2\pi i w} \frac{w}{(1+w)^2}
{\rm Tr}\, [ \, w^{J-s} {\rm e}^{-\beta H_{q}}]
\label{5-Wilson-loop-var}
\\[2mm]
&=& \int_0^\infty\frac{d\beta}{\beta}
\underbrace{\oint_{\gamma^-}
\frac{dw}{2\pi i w} \frac{w}{(1+w)^2}
{\rm Tr}\, [ \, w^{F-(p+1)} {\rm e}^{-\beta H_{q}}]}_{\mathcal{Z}_p(\beta,q)} \;. \label{tre}
\eea
where we have used different notations to be able to underline various properties.
The passage from (\ref{5-path-int}) to (\ref{5-oper-quant}) corresponds to the equivalence between
path integrals and operatorial quantization, and  $J$ and $H_{q}$ are the corresponding quantum operators
described in section \ref{sec-3}.
In (\ref{5-Wilson-loop-var}) we have employed the
Wilson loop variable $w={\rm e}^{i\phi}$, and the contour integral is along the unit circle
$|w|=1$, regulated as discussed in the last section by excluding the pole at $w=-1$.
In the last expression, eq. (\ref{tre}), we have made explicit the
fermion number operator $F=\psi\bar\psi$, as used in the Dirac index computation.
As $J=\frac12 (\psi \bar \psi - \bar \psi \psi)=
\psi \bar \psi -\frac{d}{2}$ and $s=p+1 -\frac{d}{2}$, one obtains that
$J-s=F -(p+1)$, which achieves the projection to the $(p+1,0)$-form sector.
In this last expression we have defined for convenience
the ``effective action density in proper time" $\mathcal{Z}_p(\beta,q)$
for the $(p,0)$-form gauge field with charge $q$.

Let us now analyze these formulas in various cases:

1) If susy is not gauged, the corresponding ghost term $\frac{w}{(1+w)^2}$ is absent and one obtains, setting now $s=p-\frac d2$
\begin{equation} \label{def-5}
\begin{split}
\mathcal{Z}_p^{\rm ungauged}(\beta,q) &=
\oint_\gamma \frac{dw}{2\pi i w} {\rm Tr}\, [ w^{F-p} {\rm e}^{-\beta H_q}] \\
&=
\oint_\gamma  \frac{dw}{2\pi i w}  \sum_{n=0}^d w^{n-p}\, t_n(\beta,q)\\& = t_p(\beta,q)
\end{split}
\end{equation}
where $t_n(\beta,q)$ indicates the contribution arising from the trace restricted to the Hilbert space sector
with fermion number $F=n$. No poles are present along the contour $|w|=1$,
that we indicate with $\gamma$, and the
integral extracts  from the pole at $w=0$ the contibution $t_p(\beta,q)$ due to a $p$-form.
 It  corresponds to the quantum theory
of a $(p,0)$-form  $B_{p}$  with field equations given by the twisted Dolbeault laplacian,
$(\partial_q \partial_q^\dagger + \partial_q^\dagger \partial_q) B_{p}=0$,
 satisfying no additional constraints (there is no gauge invariance for the QFT in question).
The classical equivalence $(s,q_1)\to( -s, - q_1)$  corresponds to the equivalence of the
$(p,0)$-form and  $(d-p,0)$-form effective actions, with $0\leq p\leq d$,
and with $\partial_q$ replaced by  $\partial_{\frac12-q}$.
Indeed, recalling that now $s \equiv p-\frac d2$ and $q_1 \equiv q-\frac14$, one may compute
\begin{equation}
\begin{split}
\mathcal{Z}_{d-p}^{\rm ungauged}(\beta,\tfrac12-q)&= \oint_\gamma \frac{dw}{2\pi i w} {\rm Tr}\, [ w^{J+s} {\rm e}^{-\beta H_{1/2-q}}]\\[2mm]
&=\oint_{\gamma} \frac{dw'}{2\pi i w'} {\rm Tr}\, [ {w'}^{(-J+s)} {\rm e}^{-\beta H_{q}}]\\[2mm]
&=\oint_\gamma \frac{dw}{2\pi i w} {\rm Tr}\, [ w^{J-s} {\rm e}^{-\beta H_{q}}]= \mathcal{Z}_{p}^{\rm ungauged}(\beta,q)
\end{split}
\end{equation}
where we have first written down the definition of the effective action density for the model with
couplings $(-s, -q_1)$, corresponding to $\mathcal{Z}_{d-p}^{\rm ungauged}(\beta,\tfrac12-q)$ . Then we
changed $J\to -J$ and $H_{1/2-q} \to H_{q}$, corresponding to $q_1\to-q_1$, to take into account the exchanged role of 
$(x,\psi)$ and $(\bar x,\bar \psi)$, and used $w\to w'=\frac{1}{w}$ to take into account the sign change of the gauge field $\phi\to -\phi$.
Finally, a change of variables to the original coordinate $w=\frac{1}{w'}$ shows that this expression
coincides with  the one corresponding to the couplings $(s, q_1)$.
This proves a duality between
$(p,0)$-form and  $(d-p,0)$-form at the quantum level, namely $t_{p}(\beta,q)= t_{d-p}(\beta,\tfrac12-q)$.

To check duality in our previous examples, 
it may be easier to rewrite the heat kernel coefficients in terms of the parameter $q_1\equiv q-\frac14$. For $d>1$
they read
\begin{equation}\label{non gauge p form coefficients q1}
\begin{split}
B^{(q)}_{p}
\to\ & \binom{d}{p}\times\left(1;\,-\frac{1}{12}+q_1\frac{(2p-d)}{d};\,\frac{1}{180}-\frac{p(d-p)}{24d(d-1)};\right.\\[3mm]
&-\frac{19}{1440}+\frac{p(d-p)}{24d(d-1)}+q_1\frac{(2p-d)}{12 d}+\frac{q_1^2}{6}\Big[\frac{12p(d-p)}{d(d-1)}-1\Big];\\[3mm]
&\frac{1}{288}-q_1\frac{(2p-d)}{12d}
+\frac{q_1^2}{2}\Big[1-\frac{4p(d-p)}{d(d-1)}\Big];\,
\left. -\frac{1}{240}+q_1\frac{(2p-d)}{12 d}\right)
\end{split}
\end{equation}
and for $d=1$, recalling the special format $(v_1;v_2;\bar v\equiv v_3+v_4+v_5;v_6)$, they read
\begin{equation}
\begin{split}
B^{(q)}_{p}
\to\ & \left(1;\,-\frac{1}{12} + q_1(2p-1);\, -\frac{1}{240} +\frac{q_1^2}{3};\,  
-\frac{1}{240} + \frac{q_1}{12} (2p-1)\right) \;.
\end{split}
\end{equation}

At the classical geometrical level, this duality can be understood as follows.
It is well-known that a $(p,q)$-form is Hodge dual to a $(d-q,d-p)$-form, which in turn is related
to a $(d-p,d-q)$-form  by complex conjugation. Thus a $(p,0)$-form is certainly related to a $(d-p,d)$-form.
Now, on a non-compact, topologically trivial K\"ahler manifold one may split the volume form in chiral components
using the vielbein field
\begin{equation}
g \epsilon_{\mu_1...\mu_d \bar \nu_1... \bar \nu_d} = e \epsilon_{\mu_1...\mu_d} \,
\bar e \epsilon_{\bar \nu_1...\bar \nu_d}
\end{equation}
and use  the tensor $e \epsilon_{\mu_1...\mu_d}$ to dualize the $(d-p,d)$-form to a $(d-p,0)$-form.
The correct U(1) charge assignments are seen to emerge as well, when taking care of the U(1) charge of the chiral
epsilon tensors, see appendix \ref{app:Dirac}.
 As we do  not address topological issues, this suffices for the present purposes.

2) If susy is gauged, the ghost term $\frac{w}{(1+w)^2}$ is present and one must use a prescription
to integrate over $w$. As already discussed, the correct prescription is to exclude the pole at $w=-1$.
This reproduces, in particular, the correct scalar result at $p=0$.
Duality is again obtained by $(s, q_1)\to (-s,-q_1)$, with $s=p+1-\frac d2$.
Calculating as above we obtain
\begin{equation}
\begin{split}
\mathcal{Z}_{d-p-2}(\beta,\tfrac12-q)&= \oint_{\gamma^-} \frac{dw}{2\pi i w} \frac{w}{(1+w)^2}
{\rm Tr}\, [ w^{J+s} {\rm e}^{-\beta H_{1/2-q}}]\\[2mm]
&=\oint_{\gamma^-} \frac{dw'}{2\pi i w'}
\frac{w'}{(1+w')^2}
{\rm Tr}\, [ {w'}^{(-J+s)} {\rm e}^{-\beta H_{q}}]
\\[2mm]
&=\oint_{\gamma^+} \frac{dw}{2\pi i w} \frac{w}{(1+w)^2}
{\rm Tr}\, [ w^{J-s} {\rm e}^{-\beta H_{q}}]
\\[2mm]
&=\left ( \oint_{\gamma^-} +\oint_{\gamma^0} \right )
\frac{dw}{2\pi i w} \frac{w}{(1+w)^2}
{\rm Tr}\, [ w^{J-s} {\rm e}^{-\beta H_{q}}]\\[2mm]
&
= \mathcal{Z}_{p}(\beta,q) + \mathcal{Z}^{top}_{p}(\beta,q) \;.
\end{split}
\end{equation}
Again, we have first written down the definition of the partition function at the values $(-s, -q_1)$,
then used the change of variables for the dynamical fields (the fields integrated over in the path integral)
to relate the model to its dual, thus obtaining the second line above,
where in particular $w'= e^{-i\phi}$ takes into account the sign change of the worldline U(1) gauge field.
To better interpret the resulting expression we performed a
change of integration variables $w'\to w=\frac{1}{w'}$,
which maps the regulated contour $\gamma^-$ in the $w'$ coordinates
to the contour $\gamma^+$ in the $w$ coordinates, as shown in figure 3.
\begin{figure}[h]   \centering
 \includegraphics[width=2.5in]{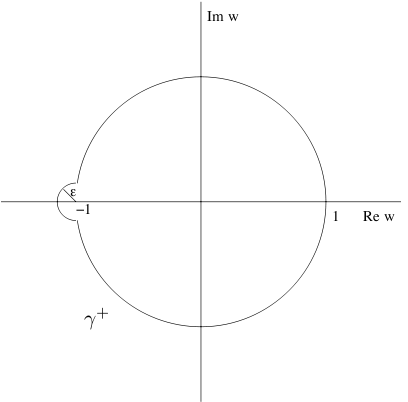}
 \caption[Fig3]     {The contour $\gamma^+$ that includes the pole at $w=-1$.}
\end{figure}

\noindent As $\gamma^+=\gamma^-+\gamma^0$, with $\gamma^0$ indicating a small contour encircling the pole at $w=-1$,
we recognize the partition function for the
gauged $(s,q_1)$ model plus a ``topological" contribution $\mathcal{Z}^{top}_p(\beta,q)$
arising form the contour integral around $\gamma^0$.
To appreciate the significance of the latter term, let us analyze it further by evaluating
the integral on $\gamma^0$ using the residue theorem
\begin{equation}\label{topological part duality}
\begin{split}
\mathcal{Z}^{top}_p(\beta,q) &=
\oint_{\gamma^0}
\frac{dw}{2\pi i w} \frac{w}{(1+w)^2}
{\rm Tr}\, [ w^{J-s} {\rm e}^{-\beta H_{q}}]
= \frac{d}{d w} {\rm Tr}\, [ w^{F-(p+1)} {\rm e}^{-\beta H_{q}}] \Big |_{w=-1}\\[2mm]
& ={\rm Tr}\, [ (F-(p+1)) (-1)^{F-p} {\rm e}^{-\beta H_{q}}]\\[2mm]
&= (-1)^p
\underbrace{{\rm Tr}\, [ F (-1)^F {\rm e}^{-\beta H_{q}}]}_{-\mathcal{Z}_{d-1}(\beta,q)}
 -(p+1)(-1)^p
 \underbrace{
 {\rm Tr}\, [(-1)^F {\rm e}^{-\beta H_{q}}]}_{{\rm ind} (\Dslash_{q-1/4})}
\end{split}
\end{equation}
The second identification in the last line in terms of the Dirac index is obvious form the discussion in section \ref{sec-3},
while the first one is proved in appendix
\ref{app:topological-form}, where it is shown that it is related to the analytic torsion of the complex manifold.

Putting all things together we obtain the following duality relation
\begin{equation}\label{duality}
\mathcal{Z}_{p}(\beta,q) =\mathcal{Z}_{d-p-2}(\beta,\tfrac12-q)
+ (-1)^{p} \mathcal{Z}_{d-1}(\beta,\tfrac12-q)
+ (-1)^{p} (p+1) \; {\rm ind} (\Dslash_{q-1/4})
\end{equation}
where we recall that the term due to a $(d-1,0)$-form  is purely topological and
carries no degrees of freedom in $d>1$.

Having found the exact duality relation \eqref{duality}, we may try to check it on some examples. 
To do so we rewrite the Seeley-DeWitt coefficients \eqref{p form coefficients} for gauge $(p,0)$-forms in terms of the parameter 
$q_1=q-\frac14$, since the duality relations are most apparent in terms of $q_1$ rather than $q$. 
As in the previous section we use the format $A_{p}^{(q)}\to(v_1;v_2;v_3;v_4;v_5;v_6)$ to present the coefficients; hence we have, 
for a gauge $(p,0)$-form in $d>3$
\begin{equation}\label{p form coefficients q1}
\begin{split}
d>3\;,&\quad0\leq p\leq d-2 \\[2mm]
A^{(q)}_{p}\to& \binom{d-2}{p}\times\left(1;\,-\frac{1}{12}-q_1\frac{d-2-2p}{d-2};\,\frac{1}{180}-\frac{p(d-p-2)}{24(d-2)(d-3)};\right.\\[3mm]
&-\frac{19}{1440}-\frac{q_1^2}{6}+(1+48q_1^2)\frac{p(d-p-2)}{24 (d-2)(d-3)}-\frac{q_1}{12}\frac{d-2-2p}{d-2};\\[3mm]
&\frac{1}{288}+\frac{q_1^2}{2}-2q_1^2\frac{p(d-p-2)}{(d-2)(d-3)}+\frac{q_1}{12}\frac{d-2-2p}{d-2};\\[3mm]
&\left.-\frac{1}{240}-\frac{q_1}{12}\frac{d-2-2p}{d-2}\right)\;.
\end{split}
\end{equation}
By noticing that, under $p\leftrightarrow d-p-2$, the number $(d-2-2p)$ goes into minus itself, it is immediate to see that 
\eqref{p form coefficients q1} is invariant under the simultaneous exchange of
 $p\leftrightarrow d-p-2$ and $q_1\leftrightarrow-q_1$, representing the duality between 
$A_p^{(q)}$ and $A_{d-p-2}^{(1/2-q)}$. The duality, as expected, does not show any topological mismatch up to order $\beta^2$ 
in $d>3$. 

On the other hand, the topological contributions are visible at order $\beta^2$ for $d\leq 3$. 
In $d=3$, the coefficients for the propagating $0$ and $1$-forms read, in terms of $q_1$,
\begin{equation}\label{d=3 01 forms q1}
\begin{split}
&d=3\;,\quad p=0,1\\[2mm]
&A^{(q)}_{p}\to\left(1;\,-\frac{1}{12}+q_1(2p-1);\,\frac{1}{180}-\frac{p}{24};\,-\frac{19}{1440}-\frac{q_1^2}{6}+
(1+48q_1^2)\frac{p}{24}+
\frac{q_1}{12}(2p-1);\right.\\[2mm]
&\left.\hspace{20mm}\frac{1}{288}+\frac{q_1^2}{2}-2q_1^2p-\frac{q_1}{12}(2p-1);\,-\frac{1}{240}+\frac{q_1}{12}(2p-1)\right)\;.
\end{split}
\end{equation}
As one can see they are not invariant under the exchange $p\leftrightarrow1-p$ and $q_1\leftrightarrow-q_1$, the difference being due to 
topological terms. To check \eqref{duality}, we compute the $v_i$ coefficients for the topological $A_2$ form
\begin{equation}\label{d=3 2 form}
\begin{split}
d=3\;,\quad p=2\;, \qquad 
A^{(q)}_{2}\to\left(0;\,0;\,\frac{1}{24};\,-\frac{1}{24}-2q_1^2;\,2q_1^2;\,0\right)
\end{split}
\end{equation}
and  can verify successfully, up to order $\beta^2$, the validity of the $d=3$ relation
\begin{equation}
\mathcal{Z}_{0}(\beta,q) =\mathcal{Z}_{1}(\beta,\tfrac12-q) +  \mathcal{Z}_{2}(\beta,\tfrac12-q) + \; {\rm ind} (\Dslash_{q-1/4})
\end{equation}
as  the Dirac index contributes only at  order $\beta^3$ (and gives a  $\beta$-independent term when inserted  in eq. 
(\ref{Seeley coefficients parametrized})).

A second nontrivial check of our duality relations may be obtained  in two complex dimensions, where the zero form
is almost selfdual
\begin{equation}
\mathcal{Z}_{0}(\beta,q) =\mathcal{Z}_{0}(\beta,\tfrac12-q) +  \mathcal{Z}_{1}(\beta,\tfrac12-q) + \; {\rm ind} (\Dslash_{q-1/4})\;.
\end{equation} 
This relation can be successfully verified by using
the scalar field coefficients, that  can be computed directly in $d=2$ from the general result (\ref{master formula U(1)}),
and seen to agree with those obtained by setting $p=0$ in \eqref{d=3 01 forms q1},
\begin{equation}\label{d=2 scalar q1}
\begin{split}
&d=2\;,\quad p=0\\[2mm]
&A^{(q)}_{0}\to\left(1;\,-\frac{1}{12}-q_1;\,\frac{1}{180};\,-\frac{19}{1440}-\frac{q_1^2}{6}-\frac{q_1}{12};\,
\frac{1}{288}+\frac{q_1^2}{2}+\frac{q_1}{12};\,-\frac{1}{240}-\frac{q_1}{12}\right)
\end{split}
\end{equation}
together with the non propagating $A_1$ form that produces the coefficients
\begin{equation}\label{d=2 1 form}
d=2\;,\quad p=1\;, \qquad 
A^{(q)}_{1}\to\left(0;\,2q_1;\,-\frac{1}{24};\,\frac{1}{24}+2q_1^2+\frac{q_1}{6};\,-2q_1^2-\frac{q_1}{6};\,\frac{q_1}{6}\right)
\end{equation}
and the twisted Dirac index of section 3 that gives
\begin{equation}
d=2\;, \qquad 
{\rm ind} (\Dslash_{q-1/4})
\to\left(0;0;\,-\frac{1}{24};\,\frac{1}{24}+2q_1^2;\,-2q_1^2;\,0\right)\;.
\end{equation}

Finally, we may have a look also at the somewhat degenerate case of $d=1$. Considering that the model at $p=-1$
is empty, the duality relation for $p=0$ collapses to 
\begin{equation}
\mathcal{Z}_{0}(\beta,q) -\mathcal{Z}_{0}(\beta,\tfrac12-q) = {\rm ind} (\Dslash_{q-1/4})
\end{equation}
that is indeed verified, after taking care of the $d=1$ relation between the  Ricci tensor and the scalar curvature,
and considering that the integral of a total derivative term  may be dropped.
Note that, for $d=1$, the $p=0$ form is not topological, but carries one degree of freedom.
This is consistent with the results in appendix \ref{app:topological-form}.

\section{Conclusions}

We have described the quantum theory of massless $(p,0)$-form gauge fields, as well as massless $(p,0)$-form fields
without gauge symmetries, using a worldline approach. The worldline  description uses a supersymmetric
nonlinear sigma model, whose backbone is the basis for proving index theorems on complex manifolds
\cite{AlvarezGaume:1983at,Friedan:1983xr}
with the physical methods of supersymmetric quantum mechanics \cite{Witten:1981nf, Witten:1982im,Witten:1982df}.
As in that case the physical motivations
for studying such models are rather indirect, as a direct spacetime interpretation is prevented by the complex nature
of the target space which allows only an even number of time directions. Nevertheless complex manifolds
find many useful applications in the context of string theory and/or  supersymmetric theories.
From a different perspective they offer a useful playground to test methods and ideas  of quantum field theory,
such as the worldline approach to theories in a curved background \cite{Bastianelli:2002fv}.
In particular, we have studied the effective action of massless $(p,0)$-forms 
on curved K\"ahler manifolds,  and discovered exact duality
relations. The calculation of several heat kernel coefficients has been presented as well.

As possible extensions of the present work one might push the calculation of the heat kernel coefficients
up to order $\beta^3$, dressing up the bosonic calculation of \cite{Bastianelli:2000dw} with fermionic contributions,
or study the duality relations  on spaces with nontrivial topology. 
Also, it could be interesting to use similar methods to study the quantum theory of $(p,q)$-forms as well as the
higher spin gauge fields introduced in \cite{Bastianelli:2009vj} on a class of complex manifolds.

\vfill\eject

\appendix
\section{Notations and conventions}
\label{app:notations}
K\"ahler manifolds can be seen as a subclass of Riemannian manifolds with additional structures.
We list here the conventions employed and some useful formulas for K\"ahler geometry, indicating occasionally their
rewriting in real coordinates, as used in Riemannian geometry.

A metric is specified by
\begin{equation}
ds^2 = G_{MN} dX^M dX^N = 2 g_{\mu\bar\nu} dx^\mu d\bar x^{\bar\nu}
\end{equation}
and the integration measure for manifolds of real dimension $D=2d$ is given by
\begin{equation}
d\mu = \sqrt{\det{G_{MN}}}\, d^D X =  \det{g_{\mu\bar\nu}}\, d^d x  d^d\bar x
\end{equation}
with the notation
\begin{equation} \label{def-measure}
  d^d x  d^d\bar x \equiv i^d \prod_{\mu=1}^d dx^\mu \wedge d\bar x^{\bar \mu} \;.
\end{equation}
For simplicity we also use the notation $ g\equiv\det{g_{\mu\bar\nu}}$.
On flat manifolds one may use cartesian coordinates for which $G_{MN}=\delta_{MN}$ and
$g_{\mu\bar\nu}=\delta_{\mu\bar\nu}$.
One can relate real and complex coordinates by
\begin{equation}
x^\mu = \frac{1}{\sqrt{2}} (X^{2\mu-1} + i X^{2\mu}) \;, \quad \bar x^{\bar \mu} = \frac{1}{\sqrt{2}} (X^{2\mu-1} - i X^{2\mu}) \;,
\quad \mu=1,...,d
\end{equation}
though other choices are also possible, of course.

We now list our conventions for connections and curvatures on K\"ahler spaces. In holomorphic coordinates the non-vanishing 
Christoffel symbols are given, in terms of the metric, by
\begin{equation}
\Gamma^\mu_{\nu\lambda}=g^{\mu\bar\mu}\de_\nu g_{\lambda\bar\mu}\;,
\quad \Gamma^{\bar\mu}_{\bar\nu\bar\lambda}=g^{\mu\bar\mu}\de_{\bar\nu} g_{\bar\lambda\mu}\;,
\end{equation}
and we shall denote their traces as
\begin{equation}
\Gamma_\mu\equiv\Gamma^\nu_{\mu\nu}=\de_\mu\ln g\;,\quad\bar\Gamma_{\bar\mu}\equiv\Gamma^{\bar\nu}_{\bar\mu\bar\nu}
=\de_{\bar\mu}\ln g\;.
\end{equation}
The non-zero components of the Riemann curvature read
\begin{equation}
R^\mu{}_{\nu\bar\sigma\lambda}=\de_{\bar\sigma}\Gamma^\mu_{\nu\lambda}\;,
\quad R^{\bar\mu}{}_{\bar\nu\sigma\bar\lambda}=\de_{\sigma}\Gamma^{\bar\mu}_{\bar\nu\bar\lambda}\;,
\end{equation}
while the Ricci tensor and the curvature scalar can be expressed as
\begin{equation}
\begin{split}
R_{\mu\bar\nu} &= -R^\lambda{}_{\lambda\bar\nu\mu}=-\de_\mu\bar\Gamma_{\bar\nu}=-\de_{\bar\nu}\Gamma_\mu
=-\de_\mu\de_{\bar\nu}\ln g\;,\\[2mm]
R &= g^{\mu\bar\nu}R_{\mu\bar\nu}\;.
\end{split}
\end{equation}
With our conventions, common in complex geometry, the curvature scalar is one half of the usual riemannian one: 
$R=\frac12 R_{(G)}\equiv\frac12 R^M{}_M$.

Let us now introduce vielbeins and spin connections, that are not used in the main text but are employed
in appendix \ref{app:Dirac} to study the Dirac operator. In holomorphic coordinates the vielbein $e_M{}^A$ splits as 
$(e_\mu^a, e_{\bar\mu}^{\bar a})$. The metric is given by $g_{\mu\bar\nu}=e_\mu^ae_{\bar\nu}^{\bar b}\delta_{a\bar b}$. 
We denote the vielbein determinants by
\begin{equation}
\det (e_\mu^a)\equiv e\;,\quad \det(e_{\bar\mu}^{\bar a})\equiv \bar e\;,
\end{equation}
so that $g=e\bar e$. Imposing the vielbein postulate $\nabla_M e_N{}^A=0$ we find for the $U(d)$ spin connection
\begin{equation}
\omega_{\mu a\bar b}=-e^{\bar\nu}_{\bar b}\,\de_\mu e_{\bar\nu a}\;,\quad \omega_{\bar\mu a\bar b}=e^{\nu}_{a}\,\de_{\bar\mu} e_{\nu\bar b}\;,
\end{equation}
while for its $U(1)$ parts we get
\begin{equation}
\omega_\mu\equiv\omega_{\mu a\bar b}\,\delta^{a\bar b}=-\de_\mu\ln\bar e\;,
\quad\bar\omega_{\bar\mu}\equiv\omega_{\bar\mu a\bar b}\,\delta^{a\bar b}=\de_{\bar\mu}\ln e\;.
\end{equation}
The Christoffel symbols are related to the spin connection via
\begin{equation}
\begin{split}
\Gamma^\mu_{\nu\lambda} &= e^\lambda_a\big(\de_\mu e_\nu^a+\omega_\mu{}^a{}_b\,e_\nu^b\big)\;,\\[2mm]
\Gamma_\mu &= -2\omega_\mu+\de_\mu\ln\frac{e}{\bar e}\;,\quad\bar\Gamma_{\bar\mu}=2\bar\omega_{\bar\mu}-\de_{\bar\mu}\ln\frac{e}{\bar e}\;.
\end{split}
\end{equation}
Finally, in order to easily compare the Seeley-DeWitt coefficients computed in the present paper with the literature, 
we list the quadratic terms in curvatures as they appear in riemannian or K\"ahler notations
\begin{equation}
R_{(G)}\equiv g^{MN}R_{MN}=2R\;,\quad R_{MN}R^{MN}=2R_{\mu\bar\nu}R^{\mu\bar\nu}\;,
\quad R_{MNRS}R^{MNRS}=4R_{\mu\bar\nu\rho\bar\sigma}R^{\mu\bar\nu\rho\bar\sigma}\;.
\end{equation}

\section{Dirac operator on K\"ahler manifolds}
\label{app:Dirac}
On K\"ahler manifolds the space of Dirac spinors is equivalent to the space
of $(p,0)$-forms with any allowed $p$, see  for example \cite{Green:1987mn}.
Here we review this decomposition and study the Dirac operator.

On real manifolds admitting spinors  it is natural to define the Dirac equation using the spin connections
$\omega_M{}^{AB}$, which is the SO($D$) connection that keeps the vielbein $e_M{}^A$ covariantly constant
\begin{equation}
\nabla_M e_N{}^A = \partial_M e_N{}^A - \Gamma_{MN}^L e_L{}^A + \omega_M{}^{AB} e_{N B} =0\;.
\end{equation}
The Dirac operator  $\Dslash\,$ is defined using the Dirac gamma matrices $\gamma^A$, which satisfy the usual Clifford algebra
$\{\gamma^A, \gamma^B\}=2\eta^{AB}$,
\begin{equation}
\Dslash = \gamma^A e_A{}^M D_M =\gamma^A e_A{}^M
\left(\partial_M +\frac14 \omega_{MBC}\gamma^B\gamma^C\right)\;.
\end{equation}

On K\"ahler manifolds one may use complex coordinates, so that curved indices split as $M\to (\mu ,\bar\mu)$,
and similarly flat indices $A \to (a ,\bar a)$. Thus the Dirac operator splits as
\begin{equation}
\label{split}
{\slashed D}=\gamma^\mu D_\mu + \gamma^{\bar \mu} D_{\bar\mu}
\end{equation}
where $\gamma^\mu= e^\mu_a \gamma^a$, $ \gamma^{\bar \mu}= e^{\bar \mu}_{\bar a}  \gamma^{\bar a}$, and
the covariant derivatives as
\begin{equation}
\begin{split}
&D_\mu =\partial_\mu +\frac12\omega_{\mu a\bar b} \gamma^a \gamma^{\bar b} -\frac12 \omega_\mu  \;,
\qquad \omega_\mu \equiv \omega_{\mu a\bar b} \delta^{a\bar b} \\
& D_{\bar\mu} = \partial_{\bar \mu} +\frac12\omega_{\bar \mu a\bar b} \gamma^a \gamma^{\bar b} -\frac12 \bar \omega_{\bar \mu}  \;,
\qquad \bar \omega_{\bar \mu} \equiv \omega_{\bar \mu a\bar b} \delta^{a\bar b}
\end{split}
\end{equation}
which shows how a precise coupling to the U(1) part of the spin connection emerges by reducing the
$SO(2d)$ connection to the $U(d)$ connection of K\"ahler manifolds.
To compare with the main text it is useful to rewrite these formulas using
the spinor variables $\psi$'s with flat tangent space indices. They are related to the gamma matrices by
$\psi^a= \frac{\gamma^a}{\sqrt{2}}$  and $\bar \psi^{\bar a}= \frac{\gamma^{\bar a}}{\sqrt{2}}$. Then the covariant derivatives take the form
\begin{equation}
D_\mu =\partial_\mu +\frac12\omega_{\mu a\bar b} \psi^a \bar \psi^{\bar b} -\frac12 \omega_\mu  \;,
\qquad
 D_{\bar\mu} = \partial_{\bar \mu} +\omega_{\bar \mu a\bar b} \psi^a \bar \psi^{\bar b} -\frac12 \bar \omega_{\bar \mu}\;.
\end{equation}

Let us now review the construction of the spinor space, i.e. the representation space of the gamma matrices.
Using the spinor variables which satisfy
 $$\{\psi^a , \bar\psi_b\}=\delta^a_b $$
one may construct
the fermionic Fock space, using $\psi^a$ as creation and $\bar\psi_a$ as destruction operators.
Thus, just as in the expansion of eq. (\ref{wave function}),
a generic spinor takes the form
\begin{equation}
\phi(x,\bar x, \psi)= F(x,\bar x) + F_a(x,\bar x)\psi^a
+\frac12 F_{a_1 a_2}(x,\bar x)\psi^{a_1}\psi^{a_2}
+ ...+\frac{1}{d!}F_{a_1..a_d}(x,\bar x)\psi^{a_1}..\psi^{a_d}\;.
\end{equation}
This shows that locally a spinor field is equivalent to the complete set of $(p,0)$-forms.

The operators $\psi^\mu D_\mu $ and $\bar \psi^{\bar \mu} D_{\bar\mu}$, obviously related to those appearing in (\ref{split}),
act on these forms as Dolbeault operators twisted by the U(1) part of the spin connection. In fact, using the vielbein to convert to tensors
with curved indices one finds
\begin{equation}
\psi^\mu D_\mu
\phi(x,\bar x, \psi)= \left (
\Big( \partial_\mu -\frac12 \omega_\mu\Big) F
\right)
\psi^\mu
+\frac12 \left(  \Big (\partial_\mu -\frac12 \omega_\mu\Big) F_\nu - \Big(\partial_\nu -\frac12 \omega_\nu\Big ) F_\mu
\right )\psi^\mu\psi^\nu+\cdots
\end{equation}
and
\begin{equation}
\bar \psi^{\bar \mu} D_{\bar\mu} \phi(x,\bar x, \psi)=  \left (
g^{\mu\bar \nu} \Big(  \partial_{\bar \nu} -\frac12 \bar \omega_{\bar \nu}\Big) F_\mu
\right)
+  \left (
 g^{\mu\bar \nu} \Big( \partial_{\bar \nu} -\frac12 \bar \omega_{\bar \nu}\Big) F_{\mu \lambda}
 \right)
 \psi^\lambda
+\cdots
\end{equation}
which contain a precise U(1) charge. Considering that in our conventions
\begin{equation} 
\omega_\mu = - \partial_\mu \ln \bar e \;, \quad
 \bar \omega_{\bar \mu}= \partial_{\bar \mu} \ln e \;, \quad
 \Gamma_\mu = \partial_\mu \ln g  \;, \quad
\bar \Gamma_{\bar\mu} = \bar \partial_{\bar \mu} \ln g
\end{equation}
with $g=\det g_{\mu\bar \nu}$, $ e=\det e_{\mu}^a$, and $\bar e= \det e_{\bar \mu}^{\bar a}$, so that $g=e \bar e$,
one finds
\begin{equation} 
\Gamma_\mu = \partial_\mu \ln g =\partial_\mu \ln e + \partial_\mu \ln \bar e =
-2 \omega_\mu +\partial_\mu \ln \frac{e}{\bar e}
\end{equation}
together with its complex conjugate expression 
$\bar \Gamma_{\bar \mu} = 2 \bar \omega_{\bar \mu} +\bar \partial_{\bar\mu} \ln \frac{\bar e}{ e}$.
These formulas allow to switch to the Christoffel connection and obtain
\begin{equation} 
\begin{split}
\psi^\mu D_\mu
\phi(x,\bar x, \psi)& = \left (
\Big (\frac{e}{\bar e}\Big )^{\frac{1}{4}}
\Big( \partial_\mu +\frac14 \Gamma_\mu\Big) \Big (\frac{\bar e}{e}\Big )^\frac{1}{4} F
\right)
\psi^\mu \\
&+\frac12 \left(  \Big (\frac{e}{\bar e}\Big )^{\frac{1}{4}} \Big (\partial_\mu +\frac14 \Gamma_\mu\Big) \Big (\frac{\bar e}{e}\Big )^\frac{1}{4}
 F_\nu - \Big (\frac{e}{\bar e}\Big )^{\frac{1}{4}} \Big(\partial_\nu +\frac14 \Gamma_\nu\Big ) \Big (\frac{\bar e}{e}\Big )^\frac{1}{4} F_\mu
\right )\psi^\mu\psi^\nu+\cdots
\end{split}
\end{equation}
and
\begin{equation} \begin{split}
\bar \psi^{\bar \mu} D_{\bar\mu} \phi(x,\bar x, \psi)&=  \left (
g^{\mu\bar \nu}
\Big (\frac{e}{\bar e}\Big )^\frac{1}{4}
\Big( \partial_{\bar \nu} -\frac14 \bar \Gamma_{\bar \nu}\Big)  \Big (\frac{\bar e}{e}\Big )^\frac{1}{4} F_\mu
\right)
\\& +  \left (
 g^{\mu\bar \nu}
 \Big (\frac{e}{\bar e}\Big )^\frac{1}{4}
 \Big( \partial_{\bar \nu} -\frac14 \bar \Gamma_{\bar \nu}\Big) \Big (\frac{\bar e}{e}\Big )^\frac{1}{4} F_{\mu \lambda}
 \right)
 \psi^\lambda
+\cdots\;.
\end{split}
\end{equation}
The $U(1)$ phase $(\frac{\bar e}{e} )^\frac{1}{4}$
can be locally eliminated by redefining the fields (or choosing a Lorentz gauge for which $e=\bar e$),
so that one may use tensor fields with curved indices and Christoffel connections only.
This proves that the Dirac operator is related to the twisted Dolbeault operators with
$U(1)$ charge $q=\frac14$, as used in the main text.
This assertion is certainly true locally, i.e. in a coordinate patch. As we do not address topological issues, apart form the use
of topological densities as found in the duality relations, this suffices for the purposes
of the present paper.

We end this appendix by reporting the U(1) charges of the chiral epsilon tensors that arise when 
splitting the volume form in chiral components  using the vielbein 
\begin{equation}
g \epsilon_{\mu_1...\mu_d \bar \nu_1... \bar \nu_d} = e \epsilon_{\mu_1...\mu_d} \,
\bar e \epsilon_{\bar \nu_1...\bar \nu_d} \;.
\end{equation}
Passing to flat tangent space indices, one may compute their covariant derivative,
which includes the spin connection, 
and check that they satisfy
\begin{equation} \begin{split}
&\nabla_\mu  \epsilon_{a_1...a_d} =  \omega_\mu \epsilon_{a_1...a_d}  \;, \quad\ \
\nabla_{\bar \mu}  \epsilon_{a_1...a_d} =  \bar \omega_{\bar \mu} \epsilon_{a_1...a_d}   \\[1mm]
&\nabla_\mu  \epsilon_{\bar a_1...\bar a_d} =  -\omega_\mu \epsilon_{\bar a_1...\bar a_d}  \;, \quad
\nabla_{\bar \mu}  \epsilon_{\bar a_1...\bar a_d} =  - \bar \omega_{\bar \mu} \epsilon_{\bar a_1...\bar a_d}   
\end{split}
\end{equation} 
as only the U(1) subgroup of the U($d$) holonomy group  does not leave the epsilon tensors invariant.

\section{Topological $\bf (d-1,0)$-form and analytic torsion}
\label{app:topological-form}

In order to find the effective action for the topological $(d-1,0)$-form in \eqref{topological part duality}, it is useful to analyze the relations among 
the effective actions of gauge $(p,0)$-forms and ``non gauge'' forms. As we have seen, they are produced by our spinning particle model 
with gauged or ungauged supersymmetry, respectively. In this appendix, we will denote with $Z_p^A(q)$ the effective action for a gauge 
$(p,0)$-form with field strength 
$F_{p+1}=\de_qA_p$, and we will refer to $Z_p^B(q)$ as to the effective action of a ``non gauge'' $(p,0)$-form obeying 
$(\de_q\de^\dagger_q+\de^\dagger_q\de_q)B_p=0$. We will extend these notations to the effective action densities as well.

In the computation of the Seeley-DeWitt coefficients of $Z^A_p(q)$ to all orders one encounters two kinds of modular integrals, 
which we shall denote $I_n(d,p)$ and $J_n(d,p)$
\begin{equation}\label{modular integrals for F}
\begin{split}
I_n(d,p) &= \int_0^{2\pi}\frac{d\phi}{2\pi}\left(2\cos\frac\phi2\right)^{d-2}e^{-is\phi}\left(\cos\frac\phi2\right)^{-2(n-1)}\\[1mm]
 &= 2^{2n-2}\oint_{\gamma_-}\frac{dw}{2\pi iw}\frac{(w+1)^{d-2n}}{w^{p+1-n}}\\[2mm]
J_n(d,p) &= \int_0^{2\pi}\frac{d\phi}{2\pi}\left(2\cos\frac\phi2\right)^{d-2}e^{-is\phi}\left(\cos\frac\phi2\right)^{-2(n-1)}\left(i\tan\frac\phi2\right)\\[1mm]
&= 2^{2n-2}\oint_{\gamma_-}\frac{dw}{2\pi iw}\frac{(w+1)^{d-2n-1}}{w^{p+1-n}}(w-1)\;,
\end{split}
\end{equation}
where $s=p+1-\frac d2$, $n\geq1$, and $w=e^{i\phi}$ is the Wilson loop variable. 
Since the regulated contour $\gamma_-$ 
excludes the pole at $w=-1$, the integrals are easily computed by the residue at $w=0$, so that
\begin{equation}
\begin{split}
I_n(d,p)&=\left\{
\begin{array}{cc}
\frac{2^{2n-2}}{(p+1-n)!}\frac{d^{p+1-n}}{dw^{p+1-n}}(1+w)^{d-2n}|_{w=0} &\quad p\geq n-1 \\
                    0 &\quad p<n-1 \\
\end{array}\right.\\[2mm]
J_n(d,p)&=\left\{
\begin{array}{cc}
\frac{2^{2n-2}}{(p+1-n)!}\frac{d^{p+1-n}}{dw^{p+1-n}}[(1+w)^{d-2n-1}(w-1)]|_{w=0} &\quad p\geq n-1 \\
                    0 &\quad p<n-1 \\
\end{array}\right.\;.
\end{split}
\end{equation}

If one wants, instead, to compute the effective action $Z^B_p(q)$ for the ungauged model, the very same heat kernel coefficients will be 
multiplied by different modular integrals $\widetilde I_n(d,p)$ and $\widetilde J_n(d,p)$, that differ from \eqref{modular integrals for F} 
by the replacement $(2\cos\frac\phi2)^{d-2}\to(2\cos\frac\phi2)^d$ and $s=p+1-\frac d2\to s=p-\frac d2$. This gives the simple identification
$$
\widetilde I_n(d,p)=4I_{n-1}(d,p-1)\;,\quad\widetilde J_n(d,p)=4J_{n-1}(d,p-1)\;.
$$
Following \cite{Bastianelli:2005vk}, we are now ready to prove the main result that will be used in deriving \eqref{duality}. For $p\geq n-1$ we have:
\begin{equation}\label{Zf Za integrals}
\begin{split}
I_n(d,p) &= \frac{2^{2n-2}}{(p+1-n)!}\de^{p+1-n}_w(1+w)^{d-2n}\Big|_{w=0}\\
&= \frac{2^{2n-2}}{(p+1-n)!}\de^{p+1-n}_w\big[(1+w)^{d-2n+2}(1+w)^{-2}\big]\Big|_{w=0}\\
&= 2^{2n-2}\sum_{k=0}^{p+1-n}\frac{\big[\de^{p+1-n-k}_w(1+w)^{d-2n+2}\big]\big[\de^k_w(1+w)^{-2}\big]}{(p+1-n-k)!k!}\Big|_{w=0}\\
&=\sum_{k=0}^{p+1-n}4I_{n-1}(d,p-1-k)(-1)^k(k+1)\;.
\end{split}
\end{equation}
Since $I_{n-1}(d,p-1-k)$ is zero for $k>p+1-n$, we can extend the sum in the above formula up to $k=p$, and recalling that 
$\widetilde I_n(d,p)=4I_{n-1}(d,p-1)$ one gets
\begin{equation}
I_n(d,p)=\sum_{k=0}^p(-1)^k(k+1)\widetilde I_n(d,p-k)\;.
\end{equation}
It is straightforward to see that an analogous formula holds as well for $J_n(d,p)$. Hence, since it holds order by order for every modular integral, 
we can conclude that it is valid for the whole effective actions, namely
\begin{equation}\label{Zf Za relation}
Z^A_p(q)=\sum_{k=0}^p(-1)^k(k+1)\,Z^B_{p-k}(q)\;.
\end{equation}
We have now all the ingredients to evaluate ${\rm Tr}[(-1)^FFe^{-\beta H_{q}}]$ in \eqref{topological part duality}
\begin{equation} \label{eatop}
\begin{split}
{\rm Tr}\Big[(-1)^FFe^{-\beta H_{q}}\Big] &= \sum_{n=0}^d(-1)^n\,n\,t_n(\beta,q)=\sum_{n=1}^d(-1)^n\,n\,t_n(\beta,q)\\
&=-\sum_{n=0}^{d-1}(-1)^n\,(n+1)\,t_{n+1}(\beta,q)
\\&=-\sum_{n=0}^{d-1}(-1)^n\,(n+1)\,t_{d-1-n}(\beta,\tfrac12-q)\\
&=-\sum_{n=0}^{d-1}(-1)^n\,(n+1)\,\mathcal{Z}^B_{d-1-n}(\beta,\tfrac12-q)=-\mathcal{Z}^A_{d-1}(\beta,\tfrac12-q)\;.
\end{split}
\end{equation}
To derive this relation we first shifted the summation variable $n$, then we used the duality for the ungauged model: 
$t_p(\beta,q)=t_{d-p}(\beta,\tfrac12 -q)$, and finally the relation \eqref{Zf Za relation} 
to recognize the effective action for the $(d-1,0)$-form. For $d>1$ this form does not carry any degree of freedom
and it is purely topological.
It can be related to the analytic torsion introduced in \cite{Ray:1973sb} 
for complex manifolds: exponentiating the effective action in (\ref{eatop})  one obtains the product of determinants of the Dolbeault laplacians
with the correct powers, as seen from the expressions in the first line of eq. (\ref{eatop}), 
which defines the analytic torsion. 

\vfill\eject

\end{document}